\documentclass[%
reprint,
groupedaddress,:
frontmatterverbose, 
showpacs,preprintnumbers,
amsmath,amssymb,
aps,
pra,
floatfix,
]{revtex4-1}

\usepackage{graphicx,url,graphics}
\usepackage{dcolumn}
\usepackage{bm}
\usepackage{longtable}                                      
\usepackage[version=3]{mhchem}
\usepackage{color}

\begin{document}
\title{Photoionization of the $3s^23p^4~ ^3P$ and the 
	$3s^23p^4~^1D,~^1S$ states of sulfur: experiment and theory}

\author{Mathias Barthel$^1$}\email{barthel@woelfel.de}
\altaffiliation{Present address: W\"olfel Beratende Ingenieure GmbH + Co. KG, Max-Planck-Str. 15,
						97204 H\"ochberg, Germany}
\author{Roman Flesch$^1$}
\author{Eckart R\"{u}hl$^1$}\email{Corresponding author: ruehl@zedat.fu-berlin.de}
\affiliation{$^1$Physikalische Chemie, Freie Universit\"{a}t Berlin, Takustr. 3, D-14195 Berlin, Germany}

\author{Brendan M. McLaughlin$^{2,3}$}\email{Corresponding author: b.mclaughlin@qub.ac.uk}
\affiliation{$^2$Centre for Theoretical Atomic, Molecular and Optical Physics (CTAMOP), 
			School of Mathematics and Physics,
			The David Bates Building, 7 College Park, 
			Queen's University of Belfast, Belfast BT7 1NN, United Kingdom\\
		$^3$Institute for Theoretical Atomic and Molecular Physics,
			Harvard Smithsonian Center for Astrophysics, 
			60 Garden Street, MS-14, Cambridge, MA 02138, USA}

%
%

\date{\today}

\begin{abstract}
Photoionization of neutral atomic sulfur in the ground and metastable states 
was studied experimentally at
a photon energy resolution of 44 meV FWHM. Relative cross section measurements 
were recorded by
using tunable vacuum ultraviolet (VUV) radiation in the energy range 9 -- 30 eV 
obtained from a laser-produced plasma and the atomic species were generated 
by photolysis of molecular precursors. Photoionization of this atom is 
characterized 
by multiple Rydberg series of autoionizing resonances superimposed on a direct 
photoionization continuum. 
A wealth of resonance features observed in the experimental spectra are 
spectroscopically assigned and their energies and 
quantum defects tabulated. The cross-section measurements are compared with 
state-of-the-art theoretical 
cross-section calculations obtained from the Dirac Coulomb {\it R}-matrix method. 
Resonances series in the spectra are identified 
and compared indicating similar features in both the theoretical and experimental spectra.
\end{abstract}

\pacs{32.80.Fb, 32.80.Zb, 32.80.Ee}

\keywords{photoionization, ions, synchrotron, radiation, resonances, metastable states}

\maketitle

\section{Introduction}
In astrophysics, abundance calculations are based on available atomic data 
that often are insufficient to make definite identifications of spectroscopic 
lines \cite{cardelli1993}. 
In planetary nebulae, the known emission lines are used to identify a
characteristic element resulting from the process of nucleosynthesis in 
stars \cite{sharpee2007,sterling2007}.
The study of the photoionization of sulfur is of considerable
interest because of its abundance in space and interstellar media.
Sulfur has been discovered in the plasma of the
Jupiter satellite Io and its occurrence in the solar atmosphere 
makes it of high astrophysical relevance \cite{Broadfoot1981,Shem1981,Symth1988}.
It is known that Io is the source of Na, K, S, and O clouds observed 
in the near-Io spatial environment and in extended regions throughout 
the Jovian magnetosphere \cite{Summers1989}.
Spectroscopic measurements indicate
that sulfur and oxygen dominate the torus plasma
of the Jupiter satellite Io. Sulfur has also been discovered in Comet Comae as
has CS with abundances of 10$^{-3}$ of water \cite{Meier1997}. 
Sulfur chemistry is also of importance 
in theoretical studies of interstellar shocks \cite{Roueff1986}. 
Photoabsorption and photoionization processes in the
vacuum ultraviolet (VUV) region play an important role in
determining solar and stellar opacities \cite{Kohl1973,Dupree1978,Kurucz1981}. 
Determining accurate abundances for atomic sulfur is of great importance in 
understanding extragalactic H II regions \cite{Garnett1989}.

The photoionization spectrum of sulfur has been studied
in detail both theoretically and experimentally. 
On the experimental side, the sulfur absorption spectrum was reported by
Tondello \cite{Tondello1972} for the energy regions below and above the 
ionization
threshold. Absolute values of the cross section for the photoionization of the 
ground state ($3s^23p^4~^3P$)
of sulfur for photon energies from the threshold at 119.67 nm to 90 nm (10.38 
-- 13.8 eV) 
using the flash-pyrolysis method, i.e. irradiating the sample with a strong 
flux of light from a discharge flash lamp. 
Sarma and Joshi \cite{Sarma1984} studied the photoionization
spectrum of sulfur using a method similar to Tondello's. 
They modified and extended Tondello's spectrum,
particularly in the region between 109 and 100 nm (11.39 -- 12.42 eV). 
Gibson {\it et al.} \cite{Gibson1986} observed the photoionization
spectrum of sulfur in the region between the first ionization
threshold and 95 nm   {(13.05~eV)}. They found that the 
autoionization
features of the $3s^23p^3 (^2D^{\circ})nd~^3D^{\circ}$ levels are broad
while that of the $3s^23p^3 (^2D^{\circ})nd~^3S^{\circ}, ^3P^{\circ}$ levels 
are narrow.
Their results are in good agreement with the measurements
of Tondello. However, they reversed the
designation given by Tondello to the $3s^23p^3 (^2D^{\circ})nd~^3S^{\circ}$ 
and $3s^23p^3 (^2D^{\circ})ns~^3D^{\circ}$ and supported these assignment
of the levels mainly on the basis of the quantum defects of the two series. 
Experimental measurements published by Joshi {\it et al.} \cite{Joshi1987} 
contained
the photoabsorption spectrum of sulfur in the wavelength
122 -- 84 nm range (10.18 -- 14.79 eV ) giving a detailed comparison of
the line list obtained in their measurement with that of
Tondello \cite{Tondello1972}, Sarma and Joshi \cite{Sarma1984}, 
and Gibson {\it et al.} \cite{Gibson1986}.
The emission spectrum of sulfur has been studied by a number
of experimental groups in the energy regions below
and above the first ionization limit. Kaufman \cite{Kaufman1979} measured
114 lines of atomic sulfur in the wavelength region 
216.9 -- 115.7 nm (5.73 eV - 10.73 eV) involving
transitions to the levels of the ground state configuration. 

Recent experiments performed by Jackson and co-workers 
\cite{Jackson2004,Jackson2005,Jackson2008a,Jackson2008b} measured the 
single-photon 
excitation spectra from the lowest $3s^23p^4~^1D_2$ and $3s^23p^4~^1S_0$ 
metastable levels of sulfur atoms
recorded with a tunable vacuum ultraviolet (VUV) radiation source generated 
by frequency tripling in noble gases. Various new lines were observed 
in the spectra which have not been previously reported. 
The photoionization efficiency (PIE) spectra of metastable sulfur (\ce{S}) atoms 
in the ($3s^23p^4~^1D_2$) and ($3s^23p^4~^1S_0$) states have been recorded 
in the 73 350 -- 84 950 $\rm cm^{-1}$ frequency range (9.094 -- 10.532 eV)
by using a velocity-map ion imaging apparatus that 
uses a tunable VUV laser as an ionization source. 

In experiments carried out by Pan and co-workers \cite{Pan2008} 
they recorded the photoionization spectra of sulfur atoms in transitions 
from the $3s^23p^4~^1D_2$ state in the energy range of
75 800 -- 89 500 $\rm cm^{-1}$ (9.4 - 11.1 eV). 
They recorded and assigned the Rydberg series
$3s^23p^3(^2D^{\circ}_{3/2})nd[3/2]$ and $3s^23p^3(^2D^{\circ}_{3/2})nd[5/2]$
with $n$ extending to 16 and 32, respectively, to the $^2D^{\circ}$ series limit.
In addition, new Rydberg series $3s^23p^3(^2D^{\circ}_{3/2})nd[1/2]$, 
$3s^23p^3(^2D^{\circ}_{5/2})nd[5/2]$ 
and $3s^23p^3(^2D^{\circ}_{3/2})nd[5/2]$ with $n$ ranging from 5 -- 9 for the 
former two series 
and 7 -- 13 for the latter were able to be assigned.

The present paper uses a technique that was applied to photoionization and 
autoionization of atomic singlet oxygen O($^{1}$D) previously \cite{Ruhl2000, 
Ruhl2008}. Suitable photochemical 
precursors of electronically excited atomic or molecular species are 
photolyzed using a pulsed, 
tunable laser. Photodissociation products are subsequently photoionized using 
a pulsed, tunable VUV light source. Investigations are facilitated in 
systems where the primary photoproduct occurs in well-defined quantum 
states. 

Carbon disulfide (\ce{CS2}) is used as a source of atomic sulfur in both the 
$^{3}P$
ground state and the first excited ($^{1}D$) metastable state. The ultraviolet
photoabsorption cross section of \ce{CS2} has been investigated previously
\cite{Leroy1981, Wu1981, Xu1993,Chen1995}. Between 180~nm 
  {(6.89~eV)} and 
220~nm   {(5.64~eV)}
there is an intense band system that corresponds to the 
$X\;^{1}\Sigma_{g}^{+} \to\; ^{1}B_{2} (^{1}\Sigma_{u}^{-})$ 
transition with photoabsorption cross
sections as high as $>$ 500~Mb for bands around 200~nm. 
The $^{1}B_{2}$ state is predissociated by 
(\ce{CS + S(^{3}P)}) and a (\ce{CS + S(^{1}D)}) continuous states. 
At 193~nm (6.424~eV) the
photoabsorption cross section is estimated to be $\sigma \approx 290\;{\rm Mb}$
using interpolated data reported in Ref. \cite{Chen1995}. 
At this wavelength photolysis can be carried out
conveniently by a pulsed ArF laser (see Section II).

Hence, atomic sulfur species S$(3s^23p^4~^3P)$ and S$(3s^23p^4~^1D)$ 
were received in  mixture by photolysis of \ce{CS2}. 
The branching ratio of both species at 193 nm   {(6.424 eV)}  has 
been the subject of several investigations. Recently, the value of the branching 
ratio of 1.66 $\pm$ 0.3 for S$(^3P)/S(^1D)$
was published \cite{Jackson2004}. The present work yields a similar value of 2.8 
$\pm$ 0.4 (c.f. ref \cite{barthel2009}), as will be discussed further below. 

Additionally, a photolytical source of \ce{S(^{3}P)} is needed in order to
separate photoionization features of \ce{S(^{3}P)} from those of \ce{S(^{1}D)}.
Disulfur monoxide (\ce{S2O}) is chosen as a reference system. Photoabsorption
and photodissociation of \ce{S2O} have been investigated in the past 
\cite{Lakshminarayana1975, Hallin1977, Chiu1982,  Clouthier1987}. In the near
ultraviolet (280-350~nm) a band system is observed that predissociates into
\ce{S(^{3}P)} and \ce{SO}. Photolysis of \ce{S2O} within this band system
is a source of atomic sulfur in only one well-defined quantum state, i.e.
$^{3}P$. Radiation from a \ce{XeCl} excimer laser ($\lambda=308\;{\rm nm}$(4.025 eV)) 
can be used to photolyze the precursor (see Section II).

On the theoretical side, photoionization cross sections calculations 
of the sulfur atom have previously been made
by Conneely {\it et al.} \cite{Conneely1970} using the close-coupling 
approximation.
These authors reported the parameters for the resonance
series in the photoionization of sulfur from the ground state. 
Mendoza and Zeippen \cite{Mendoza1988} using the {\it R}-matrix method (in the $LS$ 
coupling scheme) 
calculated the photoionization cross sections of the ground states of \ce{Si}, 
\ce{P}, and \ce{S}.
Tayal \cite{Tayal1988} using the {\it R}-matrix method (in the $LS$ coupling scheme) 
calculated the total and partial cross sections for the photoionization
of the $3s^23p^4~^3P$ ground state of the sulfur atom for photon energies from 
the first
S$^{+}(3s^23p^3~^4S^{\circ})$ ionization threshold to about 25 eV. 
Photoabsorption cross sections of the ground state of four oxygen-group atoms 
were calculated
using the eigenchannel {\it R}-matrix method by Chen and Robicheaux 
\cite{Robicheaux1994} 
within an $LS$ - $jj$ frame transformation to approximately include spin-orbit 
coupling effects. 
Tondello's \cite{Tondello1972} assignment of the 
$3s^23p^3(^2D^{\circ} )ns~^3D^{\circ}$ and
$3s^23p^3(^2D^{\circ} )nd~^3S^{\circ}$ series was reversed by Gibson {\it et 
al.} \cite{Gibson1986} based
on their considerations of the quantum defects of the two series.
Tayal \cite{Tayal1988} suggested that these two series
should not be reversed and agreed with the assignment of
Tondello \cite{Tondello1972}. 
Mendoza and Zeippen \cite{Mendoza1988} and Altun \cite{Altun1992} did
not give a conclusion about this disagreement. However, the calculations
of Chen and Robicheaux \cite{Robicheaux1994} strongly confirmed
that the assignment made by Tondello \cite{Tondello1972} was the correct one. 

The layout of this paper is as follows. In Section II we give a brief 
overview of the experimental 
procedure. In Section III we outline the theoretical methods employed in our 
work. Section IV presents 
our experimental and theoretical results. The experimental setup was 
used to determine the  
photoionization cross 
section of atomic sulfur [S$(3s^23p^4~^3P)$ and S$(3s^23p^4~^1D)$] in a broad energy range. 
  {Section} V gives a brief discussion of them. Finally in 
Section VI we give a summary of our findings.

%
%
%
%
\section{Experiment}
Single-photon-ionization in combination with time-of-flight mass spectrometry 
is used to measure state-selectively the photoion yields of atomic sulfur 
species in the gas phase. The experimental setup has been described in detail elsewhere 
\cite{Flesch2000}.
Briefly, the experimental setup consists essentially of the following components:

\begin{enumerate}

\item A tunable, pulsed VUV light source based on a laser-produced
plasma \cite{Turcu1998}.
The plasma is generated by focusing a Nd:YAG-Laser 
(Spectron Laser Systems, SL 400; 1064 nm, 500 mJ/pulse, 10 Hz, 6 ns pulse 
length) 
onto a tungsten wire target. A glass capillary is used to transfer the plasma 
radiation on the entrance slit of a normal incidence monochromator 
(McPherson, Model Nova 225). Plasma radiation is dispersed by a 
gold coated spherical reflection grating (1200 lines/mm). 
The absolute number of  VUV photons  is estimated to be of the 
order of $10^8$ photons per pulse, corresponding to 
$10^9$ photons  per second at 10 Hz operation of the lasers, which is similar 
to previous work
\cite{Ruhl2000, Ruhl2001,Ruhl2002,Ruhl2004,Ruhl2008}. The typical bandwidth
of the dispersed VUV photons is set to $\Delta\lambda~\approx$ 2.4 - 5.9 $\AA$, 
depending on the spectral features to be resolved in the respective energy 
regime. 
Calibration of the VUV source and its photon energy scale is performed by 
taking ion yields of molecular oxygen \cite{Flesch2000, Berkowitz1979}. The 
well-known vibrationally resolved  transitions in the energy range 
between 12.19 and 13.78 eV are used for  this purpose \cite{Berg1991}. 

\item Pulsed ultraviolet lasers are used to photolyze precursor molecules. An 
\ce{ArF} excimer laser (Lambda Physics, OptexPro, $\lambda=193\;{\rm nm}$, 
10~mJ/pulse, 10~Hz, 10~ns pulse length)
serves to photolyze   {\ce{CS2}}. 
The pulse energy is reduced to approximately 50 $\mu$J due to the occurrence 
of strong  multi-photon ionization signals from the photolysis laser. 
\ce{S2O} photoexcitation is carried out by a  \ce{XeCl} Laser (Lambda Physik LPX 202i;
200~mJ/Pulse, 10~Hz, 10~ns pulse length). The laser radiation is not attenuated
since multiphoton processes are not observed for \ce{S2O}.

The photolysis laser is time-correlated with the VUV-system by an external
pulse generator. The delay time between
the 193 nm   {(6.424 eV)} pulse and the VUV pulse in of the order of 100--400~ns.

\item A time-of-flight mass spectrometer is used to detect cations resulting 
from photoionization by tunable VUV-radiation. The mass 
spectrometer works according to the Wiley-McLaren energy- and space
focusing conditions
with a typical mass resolution of $m/\Delta m \approx 100$ 
\cite{Wiley1955,Ruhl1991}.
Typical pressures in the ionization region of the time-of-flight mass 
spectrometer are of the order
of $5 \cdot 10^{-6}\;{\rm mbar}$.

\end{enumerate}
\ce{CS2} is used in commercial quality (Sigma Aldrich) without further
purification. It is effusively introduced into the high vacuum chamber by a
needle valve. 

Disulfur monoxide (\ce{S2O}) is synthesized 
using the method of Schenk and
Steudel \cite{Schenk1966,Schenk1964a,Schenk1964b}: 

\ce{SOCl2 + Ag2S $\rightarrow$ S2O + 2 AgCl}.

\noindent
\ce{S2O} is introduced into the ionization region in a mixture with rare gases,
such as argon or helium, which is due to the synthesis procedure.

Pump-probe photoionization mass spectra are recorded at constant VUV-photon energy. 
Photoion yields of specific masses, such as $m/z=32$ (sulfur) are obtained
from selecting the corresponding time-of-flight signal in a set of mass spectra
that are recorded as a function of the VUV-photon energy.

Ion yields are normalized with respect to the photon flux by using a
photomultiplier that is coated by sodium salicylate as a quantum converter
\cite{Samson1967}.
%
%
%
%
%
\section{Theory}
The study of the photo-absorption spectrum of atomic sulfur is interesting
due to the open-shell features of this atom and the role played by
electron correlation effects. In order to gauge the quality of our theoretical work we performed 
large-scale close-coupling calculations and compared them to the present
experimental measurements. We have used an extended large-scale close-coupling 
model in our calculations on this system compared to our previous preliminary Breit-Pauli calculations \cite{barthel2009}.
In the present theoretical work we include 512 levels of the residual \ce{S^+} ion in the close-coupling calculations 
and perform all the cross section calculations within the relativistic Dirac Coulomb {\it R}-matrix approximation.
%
\begin {figure} [htp]
 \vspace{.2in}
 \includegraphics[width=3in]{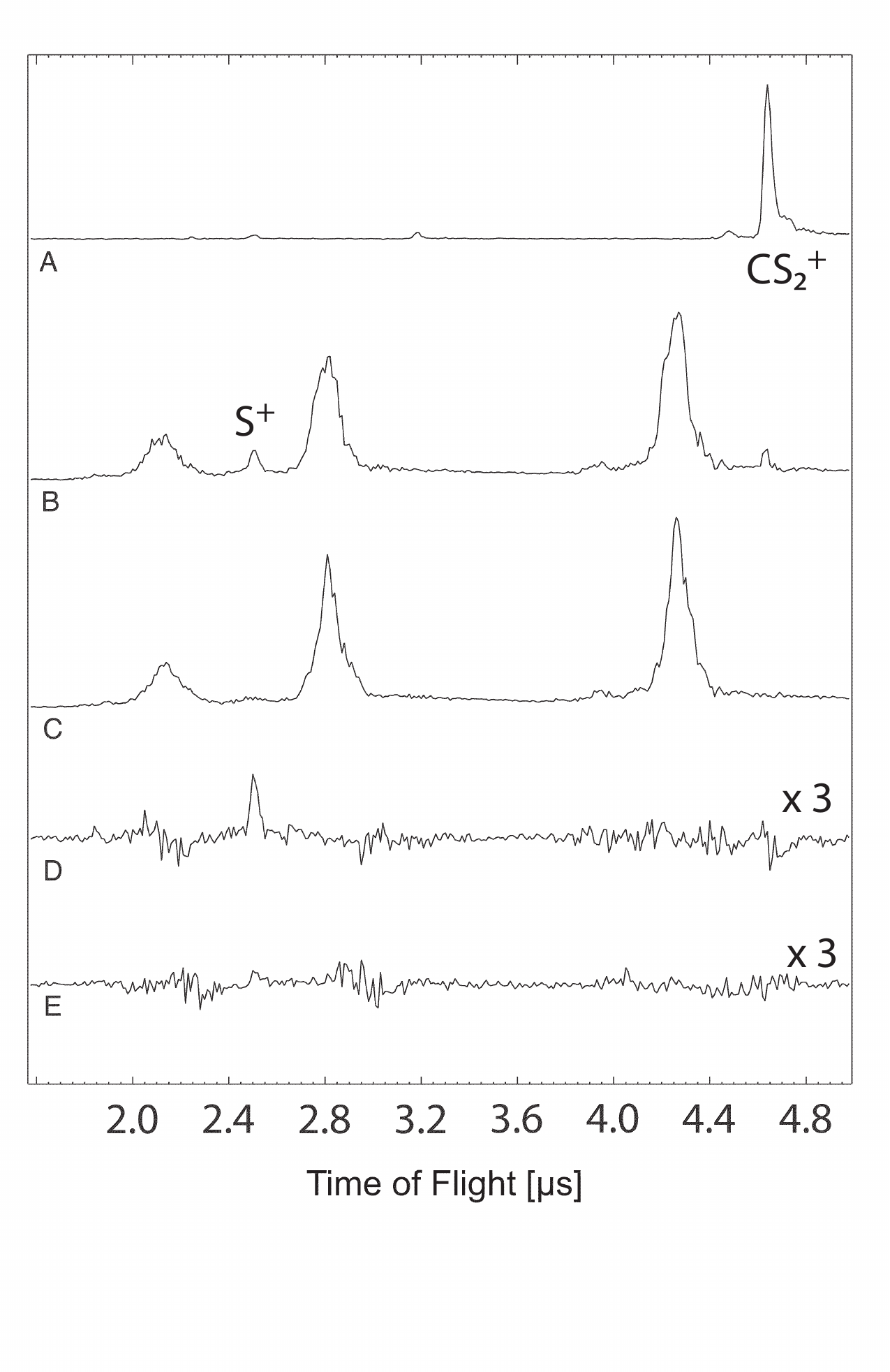}
\vspace{.2in}
\caption { Time-of-flight mass spectra of carbon disulfide 
		(\ce{CS2}) recorded at different excitation conditions. (A) photoionization at 
		$\lambda$=115.87~nm (10.70~eV); (B) primary photoexcitation 193~nm    {(6.424~eV)}
		and subsequent photoionization (10.70~eV); (C) photoexcitation 193~nm    {(6.424~eV)}
		without subsequent photoionization, indicating multi-photon ionization 
		caused by 193 nm   {(6.424 eV)} radiation; (D) difference (B)-(C); (E) similar to (D), but 
		photoionization at 10.25~eV. \label {picTOFMS}}
\end {figure}

%
%
%
\begin{figure*}
\begin{center}
\includegraphics[scale=1.5,width=\textwidth]{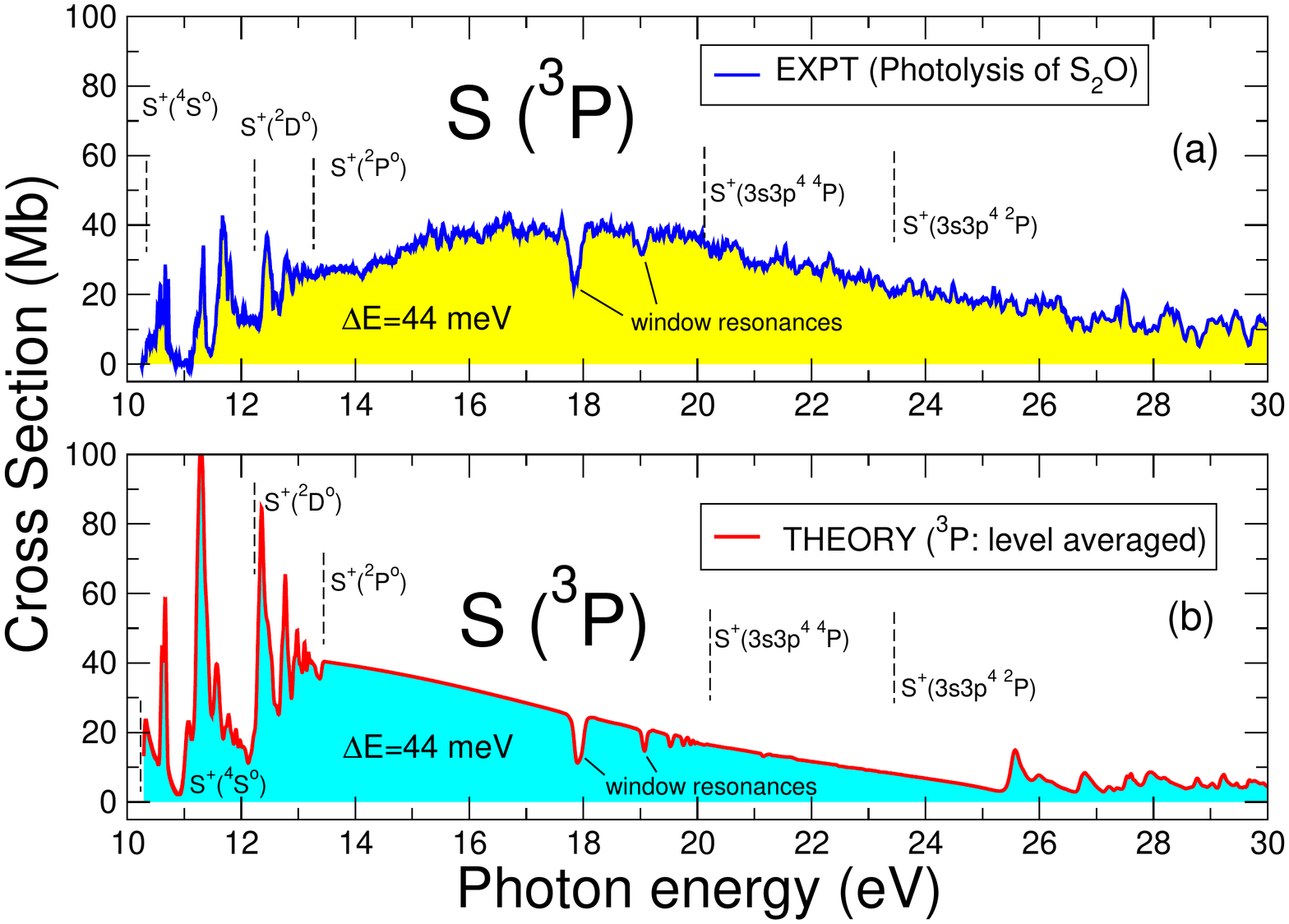}
\caption{\label{3P}  (Color online) Photoionization of S$(3s^23p^4~^3P)$ in the energy range from threshold up to 30 eV.
The various Rydberg series limits are illustrated, namely
S$^+[3s^23p^3 (^2D^{\circ}_{3/2,5/2})]$ and S$^+[3s^23p^3 (^2P^{\circ}_{1/2,3/2})]$ 
by vertical dashed lines. The prominent window resonances are clearly visible. (a) The experimental 
cross section data are taken at a nominal energy resolution of 44 meV. (b) Theoretical cross sections 
for   {100\% of} the S$(3s^23p^4~^3P)$ initial state from the 512-level DARC calculations (level averaged), convoluted 
with a Gaussian profile having a FWHM of 44 meV.
The corresponding series limits $E_{\infty}$ of Equation (2) for each series are 
indicated by vertical-dashed lines. }
\end{center}
\end{figure*}

\subsection{Dirac-Coulomb {\it R}-matrix}
Recently developed Dirac Atomic {\it R}-matrix Codes (DARC) for parallel computing architectures
\cite{norrington87,norrington91,darc,ballance06,Ballance2012,McLaughlin2012} were used
to treat photon interactions with this neutral atomic system. This suite of collision codes has the capability to 
cater for hundreds of levels and thousands of scattering channels 
\cite{venesa2012,Ballance2012,McLaughlin2012}.
Metastable states are populated in the present sulfur atom experiments and require
additional theoretical calculations. We note that our work is 
of prime interest to astrophysics as outlined in the Introduction. 
  {High quality photoionization cross section calculations have 
been made on several complex systems (Fe-peak elements and Mid-Z atoms) 
of prime interest to astrophysics and plasma applications,   
indicating suitable agreement with high resolution measurements made at leading synchrotron light sources 
\cite{Ballance2012,McLaughlin2012,Hino2012,Mueller2014,Kennedy2014}}.

To benchmark theoretical results with the experimental measurements,
photoionization cross sections calculations on this sulfur atom were 
performed for both the ground and the excited metastable levels 
associated with the $3s^23p^4$ configuration. 
Hibbert and co-workers have shown that two-electron 
promotions are important to include to get accurate 
energies, $f$-values and Einstein coefficients \cite{Ohja1989,Keenan1993} which 
are included in the present study.
In our photoionization cross-section calculations for 
this element, all 512 levels arising from the eight configurations: 
$3s^23p^3$, $3s3p^4$, $3s^23p^23d$, $3s^23p3d^2$,
$3s3p^33d$, $3p^33d^2$, $3p^5$, $3s3p^23d^2$ of the residual sulfur singly ionized ion
were included in the close-coupling expansion.  
PI cross section calculations with this 512-level model were performed in 
the Dirac Coulomb approximation using the recently developed parallel version of the DARC 
codes \cite{Ballance2012,McLaughlin2012}. 

The {\it R}-matrix boundary radius of 13.28 Bohr radii was sufficient to envelop
the radial extent of all the $n$ = 3 atomic orbitals of the residual S$^{+}$ ion. 
A basis of 16 continuum orbitals was sufficient to span the incident experimental photon energy
range from threshold up to 50 eV. This resulted in generating a maximum of 2,696 coupled 
channels in the close-coupling calculations with dipole and Hamiltonian 
matrices of the order of $\sim$ 32,525 in size. Similarly, here for the
ground-state configuration, photoionization out of the $3s^23p^4~^3P_{2,1,0}$ levels require the 
bound-free dipole matrices, $J^{\pi}=2,1,0^{e} \rightarrow J^{\pi}=0^{\circ}, 1^{\circ}, 2^{\circ}, 3^{\circ}$ 
and for the excited $3s^23p^4~^1D_2$ and $3s^23p^4~^1S_0$ metastable states, 
the bound-free dipole matrices, $J^{\pi}=0^{e}, 2^{e} \rightarrow J^{\pi}=1^{\circ}, 2^{\circ}, 3^{\circ}$. 

The current state-of-the-art parallel DARC codes running on high performance computers (HPC)
world-wide, allows one to concurrently form and diagonalize 
large-scale Hamiltonian and dipole matrices \cite{McLaughlin2014a,McLaughlin2014b} required for
electron or photon collisions with atomic systems. This allows large-scale computations to be 
completed in a timely manner.

\subsection{Photoionization}
In our calculations for the ground and metastable levels, the outer region electron-ion collision 
problem was solved with a fine mesh of 5$\times$10$^{-7}$ Rydbergs ($\approx$ 6.8 $\mu$eV) 
to fully resolve the extremely narrow resonance features in the appropriate photoionization cross sections. 
The $jj$-coupled Hamiltonian diagonal matrices were adjusted so that the theoretical term
energies matched the recommended experimental values of NIST \cite{nist}. We note that this energy
adjustment ensures more reliable positioning of the resonances relative to all thresholds included in
the calculations. Finally, in order to compare with experimental measurements, the theoretical 
cross-section calculations were convoluted with a Gaussian having 
a profile of width similar to the experiment resolution (44 meV FWHM).

\subsection{Resonance structure}
The energy levels tabulated from 
references \cite{sugar1991,saloman2007} 
and from the NIST tabulations \cite{nist} were
 used as a helpful guide for the present assignments. 

The resonance series identification can be made from Rydberg's formula:
\begin{equation}
\label{rydberg}
\epsilon_n = \epsilon_{\infty} - \frac{{\cal~Z}^2} { \nu^{2}}  
\end{equation} 
\noindent
where in Rydbergs $\epsilon_n$ is the transition energy, 
$\epsilon_{\infty}$ is the ionization potential of the excited electron 
to the corresponding final state ($n = \infty$), i.e. the resonance 
series limit \cite{Seaton1983} and $n$ being the principal quantum number.
The relationship between the principal quantum number $n$, 
the effective quantum number $\nu$ and the quantum defect $\mu$ 
for an ion of effective charge ${\cal Z}$ is given by $\nu$ = $\rm n - \mu$.
Converting all quantities to eV we can represent the Rydberg resonance series as;
\begin{equation}\label{eV}
E_n = E_{\infty} - \frac{{\cal{Z}}^2 {\rm R}}{(n - \mu)^2} .
\end{equation} 
\noindent
Here, $E_n$ is the resonance energy, $E_{\infty}$ the resonance series limit, 
$\cal{Z}$ is the charge of the core (in this case $\cal{Z}$ = 1), $\mu$ is the quantum defect, being zero for a pure
hydrogenic state, and the Rydberg constant R is 13.6057 eV. 

The multi-channel {\it R}-matrix eigenphase derivative (QB) technique, which is 
applicable to atomic and molecular complexes, of Berrington and 
co-workers \cite{qb1,qb2,qb3} was used to locate and determine the 
resonance positions in Tables \ref{meta-1D} and \ref{meta-1S}. 
The resonance width $\Gamma$ is determined from 
the inverse of the energy derivative of the eigenphase sum $\delta$ at the resonance energy $E_r$ via
\begin{equation}
\Gamma = 2\left[{\frac{d\delta}{dE}}\right]^{-1}_{E=E_r} = 2 [\delta^{\prime}]^{-1}_{E=E_r} \quad.
\end{equation}
%
%
%
%
\begin{figure*}
\begin{center}
\includegraphics[scale=1.5,width=\textwidth]{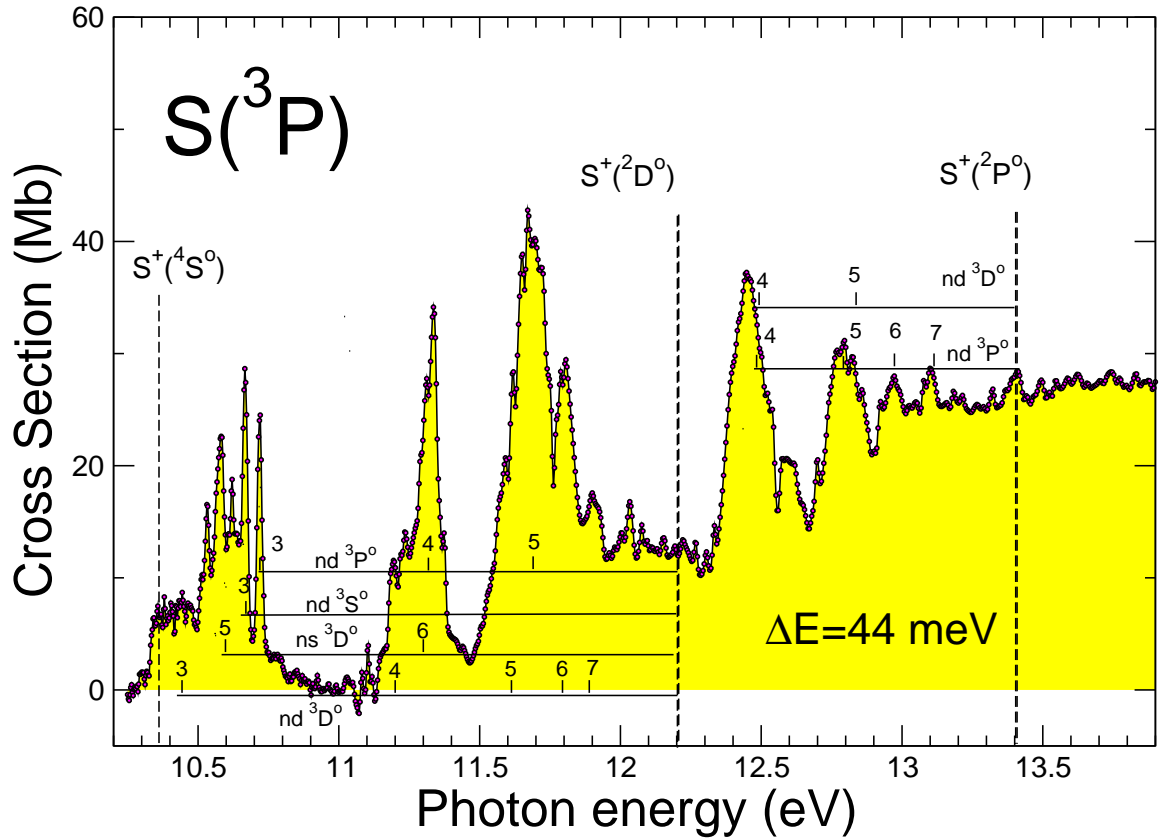}
\caption{\label{thres3P}  (Color online) An overview of the present single photoionization cross-sections 
measurements as a function of the photon energy below 13.5 eV. 
The experimental measurements were made at a nominal energy resolution of 44 meV. 
The assigned Rydberg series limits are indicated as vertical lines grouped by horizontal lines. 
The corresponding series limits $E_{\infty}$ of Equation \ref{eV} for each series are 
indicated by vertical-dashed lines in the end of the line groups. See text for 
a discussion of these resonance features. } 
\end{center}
\end{figure*}

%


\section{Results}

\subsection {Experimental Results}
Fig. \ref{picTOFMS} shows time-of-flight mass spectra of carbon disulfide 
(\ce{CS2}) recorded at different excitation conditions. The top 
spectrum (Fig. \ref{picTOFMS}(A)) shows the photoionization of \ce{CS2} where 10.70~eV radiation is 
used. This photon 
energy is well above the photoionization threshold of
10.076~eV of \ce{CS2} \cite{Coppens1979}. The parent 
cation \ce{CS2^{+}} ($m/z=76$)  is observed at t=4.64~$\mu$s. Additional weak intensity 
above t=4.64~$\mu$s comes from isotopomers such as \ce{C^{32}S^{34}S}. Weak signals 
occur at t=2.51~$\mu$s ($m/z=32$ ) and at t=3.18~$\mu$s ($m/z=44$). 
They are assigned to \ce{S^{+}} and \ce{CS^{+}} , respectively.
These  do not come from 10.70~eV photoionization of \ce{CS2} since the 
threshold of ionic fragmentation of \ce{CS2} is 14.80~eV \cite{Coppens1979}.
Rather, these ions are due to dissociative photoionization of \ce{CS2} that is caused by 
second-order radiation from the monochromator (E= 21.4~eV, see Section II.). A weak
signal at $t=4.48\;\mu$s is attributed to impurities in the sample.

Fig. \ref{picTOFMS}(B) shows the mass spectrum for primary photolysis of \ce{CS2}
(193~nm   {(6.424 eV)}) with subsequent ($\Delta t$=400~ns) photoionization at 10.70~eV. This 
spectrum is drawn on the same scale as Fig. \ref{picTOFMS} (A). The 
\ce{CS2^{+}}-signal is much weaker, whereas additional broad signals are 
observed around 2.1~$\mu$s, 2.8~$\mu$s, and 4.25~$\mu$s, respectively. These are due 
to dissociative multiphoton ionization of \ce{CS2}. They are time-shifted 
by $\approx$ 0.4~$\mu$s, which is the delay time between the 193-nm photolysis and subsequent 
photoionization by a 10.70~eV photon pulse. This delay facilitates the assignment of 
signals that come from multiphoton ionization. The width of these signals 
is due to (i) a jitter in the delay time between the laser pulses and (ii) 
a kinetic energy release during the multiphoton ionization process that 
is not discussed further. 

The main feature of mass spectrum (B) regarding the formation of atomic sulfur by photolysis
is an enhanced intensity of the \ce{S^{+}} signal at $t=$2.51~$\mu$s. 
This is apparently {\em not} due to 
multiphoton ionization since the signal is absent if {\em exclusively} 193-nm 
radiation is used (cf. Fig. \ref{picTOFMS} (C)). 

The difference between the spectra \ref{picTOFMS} (C) and (B) is shown in 
Fig. \ref{picTOFMS} (D). The only signal that remains after the 
subtraction is \ce{S^{+}}.  It implies that the 
\ce{S^{+}} signal comes from a two-step process:

\noindent
(i) photolysis of neutral \ce{CS2} (193~nm   {(6.424 eV)} radiation), leading to the formation of atomic sulfur according to

\ce{CS2 ->[193~\rm{nm}][] S + CS}

\noindent
(ii) photoionization of atomic sulfur generated in step (i):

\ce{S ->[10.70~{\rm eV}][] S^{+}}.

The ionization energy of atomic sulfur in its ground state S$(^{3}P_{2})$, 
leading to \ce{S^{+}}$\left(^{4}S^{0}_{3/2}\right)$ is 10.36001~eV \cite{Martin1990}, so 
that \ce{S^{+}} from photoionization of S($^{3}P_{2}$) is a source of 
\ce{S^{+}}. Electronically excited  \ce{S(^{1}D2)} has an excitation energy of 
1.145~eV \cite{Martin1990}   {so that the ionization energy 
of this state is reduced to 9.215~eV. Photoionization of the   \ce{S(^{1}D)}
state into the $^{4}S$ continuum is possible from an energetic point of view; however, }
 \ce{S(^{1}D)} can only  be autoionized at this photon energy, since direct photoionization into the
$^{4}S^o$ ground state of \ce{S^{+}} is spin-forbidden. 

The \ce{S^{+}} signal may therefore contain contributions from both \ce{S(^{3}P)}
and \ce{S(^{1}D)}.  Selecting this signal and recording its intensity 
as a function of the VUV energy leads to the \ce{S^{+}} ion yield of 
photolyzed \ce{CS2} which may come from both quantum states produced in \ce{CS2} 
photolysis, i.e.  \ce{S(^{3}P)} and  \ce{S(^{1}D)}. There are two difficulties 
connected with the \ce{S^{+}} ion yield from 193 nm   {(6.424 eV)} 
photolysis of \ce{CS2}:

(i) 
Atomic sulfur is produced in a mixture of it lowest-lying $^{3}P$ and 
  {$^{1}D$} states, as outlined in Section I. 
In order to separate the $^{1}D$ yield from the $^{3}P$ 
yield, \ce{S2O} is used as a reference system, which exclusively yields  \ce{S(^{3}P)}  upon photolysis. 

The experimental procedure is similar to that described above for \ce{CS2}. Photolysis 
and subsequent photoionization yields \ce{S^{+}}. By recording the \ce{S^{+}} 
intensity as a function of VUV photon energy we obtain the yield for 
photoionization and autoionization of pure \ce{S(^{3}P)}.

The \ce{S^{+}} ion yield from \ce{S2O} is scaled with respect to the \ce{S^{+}} 
yield from \ce{CS2} so that the relative intensities of resonant features that 
come from photoioization of \ce{S(^{3}P)} have the same intensity in both spectra. The 
\ce{S^{+}} yield originating from \ce{S2O} is subsequently subtracted from the \ce{S^{+}} yield 
which is obtained from \ce{CS2}. This provides the ionization yield of pure \ce{S(^{1}D)}, provided that other 
sources of \ce{S^{+}} are of negligible importance. 

(ii) 
We note that there are indeed additional channels at higher VUV photon energy 
leading to the formation of \ce{S^{+}}, so that the \ce{S^{+}} yield may have other origins than \ce{S(^{1}D)}. 
These are listed in Table \ref{tabAppErg}, where channels (3) and (4) from \ce{CS2} as well as (6) and (7) from 
\ce{S2O} need to be considered.

As a result, additional subtraction processes have to be carried out in order 
to remove such additional \ce{S^{+}} intensity, originating from other sources besides \ce{S(^{1}D)}, 
as indicated in Table \ref{tabAppErg}. 
Specifically, the \ce{S^{+}} yield that is attributed to ionization of 
\ce{S(^{1}D)} likely contain contributions from such channels, as will also be outlined in the following 
in comparison with the modeling results. 

\begin {table*}[htbp]
 \footnotesize
\centering
\caption{Threshold energy (TE) in eV of \ce{S+} from \ce{CS2} and \ce{S2O} precursors. The product channels
		are sorted by increasing threshold energy (eV) for each precursor. Secondary
		photodissociation products of primary photofragments, such as S from CS or SO
		photodissociation with subsequent photoionization are
		considered to be of minor intensity, so that they are not listed. Calculated threshold energies (eV) are based on
		data taken from Ref. \cite{Darwent1970}. Transition probabilities and selection rules are not taken into account.}
\label{tabAppErg}
\renewcommand{\arraystretch}{1.5}
\begin{center}
\begin{tabular}{m{1cm}m{1cm}m{10cm}m{1cm}} \hline\hline
Source & $\frac{\rm TE}{\rm (eV)}$ & Pathway to \ce{S+} & \\ \hline
\ce{CS2} & ~~9.21 & \ce{CS2 ->[\ce{photolysis}][]} \ce{S($^{1}$D)}+\ce{CS} \ce{->[VUV]} \ce{S+} + \ce{CS} &(1) \\
\ce{CS2} & 10.36 & \ce{CS2 ->[\ce{photolysis}][]} \ce{S($^{3}$P)}+\ce{CS} \ce{->[VUV]} \ce{S+} + \ce{CS} &(2) \\
\ce{CS2} & 14.80 & \ce{CS2 ->[\ce{VUV}][]} \ce{CS} \ce{+} \ce{S+} &(3)  \\
\ce{CS2} & 18.21 & \ce{CS2 ->[\ce{photolysis}][]} \ce{S($^{3}$P}, $^{1}$D)\ce{+ CS} \ce{->[VUV]} \ce{S} + \ce{CS+} \ce{->[{\textrm {ionic}}][\textrm{fragm.}] S + C + S+} &(4) \\ \hline
\ce{S2O} & 10.36 & \ce{S2O ->[\ce{photolysis}][]} \ce{S($^{3}$P)}+\ce{SO} \ce{->[VUV]} \ce{S+} + \ce{SO} &(5) \\ 
\ce{S2O} & 13.76 & \ce{S2O ->[\ce{VUV}][]} \ce{SO} \ce{+} \ce{S+} &(6) \\
\ce{S2O} & 15.72 & \ce{S2O ->[\ce{photolysis}][]} \ce{S($^{3}$P)}+\ce{SO} \ce{->[VUV]} \ce{S} + \ce{SO+} \ce{->[{\textrm {ionic}}][\textrm{fragm.}] S + S+ + O} &(7) \\ \hline
\hline 
\end{tabular}
\end{center}
\raggedright
\begin{flushleft}
\vspace{2mm}
Notes: (1-2, 5): taken from Ref. \cite{Martin1990}; (3): cf. Ref. \cite{Coppens1979};  (4, 7): 
			calculated by using the dissociation energy of \ce{CS} and \ce{SO} according 
			to Ref. \cite{Darwent1970} and the ionization energy of \ce{S}$\left(^{3}P\right)$ 
			according to Ref. \cite{Martin1990}; (6): taken from Ref. \cite{barthel2009}.
\end{flushleft}
\end{table*}

\normalsize

\subsection{Cross-Sections}

The electronic ground state of atomic sulfur is $(3s^23p^4~^3P)$ having the electron configuration
$3s^23p^4$ and by spin-orbit effects splits into the three fine-structure states namely, $3s^23p^4~^3P_{2,1,0}$. 
The difference in energy between the ground state $3s^23p^4~^3P_2$,
$3s^23p^4~^3P_1$, and the $3s^23p^4~^3P_0$ states is 49 and 71 meV \cite{nist}, respectively. 
Note, at room temperature the sulfur atom is primarily in the $3s^23p^4~^3P_2$ state, 
which is why contributions from the other two states may be neglected.
Similarly, the electronic configuration of the positively charged sulfur ion (S$^+$) 
in the $3s^23p^3$ configuration can form the states $^4S^o$, $^2D^o$, and $^2P^o$. 
The continuum states are energetically located at 
10.36 eV, 12.20 eV, and 13.40 eV above the electronic
ground state of the neutral atom \cite{nist}. The states formed are
analogous to isoelectronic atomic oxygen, taking into account
the emitted photoelectron terms that are relevant to the optical transitions. 
The uni-positive $^4S^o$ state with spin $S =3/2$, together with
the electron spin $s$ = $\pm$ 1/2 gives a total spin of 1 or 2, hence
triplet and quintet terms may arise. An analogous procedure results for
the case of the $^2D^o$ and $^2P^o$ excited states, 
where both singlet and triplet terms are formed.
Optical excitation from the ground state of the $3s^23p^4~^3P_2$
sulfur atom in the sense of the spin selection rule $\Delta ~S = 0$ then
makes all three continuum states allowed. 

Few experimental measurements have been performed
for the absolute photoionization cross section 
of the ground state of atomic sulfur. As pointed out in the Introduction,
Tondello \cite{Tondello1972} measured the absorption spectrum of atomic 
sulfur giving the first values for the ionization cross section. 
Absolute photoionization cross sections
at a photon energy of 14.76 eV were published by Innocenti et al. \cite{Innocenti2007} 
who found a cross section value of 75 Mb. Extrapolation of these results to a photon energy of 16.7 eV, gave 
a value of 68.7 Mb having a 50\% absolute error and larger than the 50 Mb estimated from the 
absolute measurements of Joshi and co-workers \cite{Joshi1987} which are estimated to have a 25\% error. 
This photon energy was chosen as theoretical work is available
from the Hartree-Fock-Slater method by Yeh and Lindau \cite{Yeh1985} 
which gave a cross-section value of 18.2 Mb. 
The accuracy of the measurements carried out by Tondello \cite{Tondello1972} 
are 50\% higher than these very approximate calculations.
In the present work the absolute photoionization
cross section for S$(^3P)$ at a photon energy of 16.7 eV
yields a value of 43.5 $\pm$ 10 Mb. 
Preliminary calculations using the Breit-Pauli approximation (in the framework of the {\it R}-matrix method
at this photon energy) produces a value of 42.8 Mb \cite{barthel2009}. 
The present level averaged DARC calculations produces a value of 33 Mb. These theoretical values 
are in accord with both the {\it R}-matrix $LS$-coupling results of Tayal \cite{Tayal1988} 
and the many-bodied perturbation estimates carried our by Altun \cite{Altun1992}. 
The present photoionization cross-section calculations are greatly extended 
within the confines of the Dirac {\it R}-matrix method. We investigated the photon energy region from 
thresholds up to 30 eV. All the resonance structure converging 
to the S$^+(^2D^o_{3/2,5/2})$ and S$^+(^2P^o_{1/2, 3/2})$ thresholds
and the associated window resonances converging to the  S$^+(3s3p^4~^{2,4}P_J)$ 
ionic thresholds is analyzed and discussed. 

\subsection{Photoionization of S$(3s^23p^4~^3P)$}

Photoionization of the ground state of atomic sulfur S$(3s^23p^4~^3P)$ was examined 
using the photodissociation of \ce{S2O}, which is initiated by photolysis with 308 nm (4.025 eV) radiation, yielding \ce{S(^{3}P)}. Detection
is accomplished by one-photon ionization using VUV radiation from a laser produced plasma. Using the
observed autoionization resonances in the photon energy range below 13.40 eV and the window 
resonances converging to the $^4P$ threshold at 20.20 eV, allows us to identify S$(3s^23p^4~^3P)$ as the sole photolysis
product from \ce{S2O}. The experimental technique of the present work is a
valuable alternative to that used by other authors who used \ce{H_2S} with 
hydrogen and fluorine radicals in a flow system \cite{Gibson1986,Berkowitz1994}.

An overview of the experimental photoionization spectrum from the S$(3s^23p^4~^3P)$ ground state taken at an energy resolution 
of $\Delta E$=44 meV is show in Figure \ref{3P} (a) and the results from the DARC calculations 
are illustrated in Figure \ref{3P} (b). The theoretical cross sections 
were convoluted with a Gaussian having a width of 44 meV 
in order to simulate the experimental results. We note that strong resonance features are observed
in the cross section below 13 eV and there is a prominent window resonance located at approximately 18 eV.
Figure \ref{thres3P} illustrates the experimental cross section for the S$(3s^23p^4~^3P)$
ground state in the photon energy region 
from threshold to approximately 13.9 eV. The intensity of the resonant structures 
  {in the cross sections below 12 eV in}
comparison with the continuous region 14 -- 17 eV appear to be more prominent in the theoretical spectrum 
than in the experimental results, due to the limited energy resolution available in the experimental results. 
In addition to the direct photoionization, there are a number of partial auto-ionizing Rydberg series that are optically allowed. 
We note, from the S$(3s^23p^4~^3P)$ ground state of atomic sulfur, excitation into the $ns$ 
and $nd$ Rydberg states are possible giving the $3s^23p^3(^2D^o)n\ell~^3L^o$ and $3s^23p^3(^2P^o)n\ell~^3L^o$ Rydberg 
resonance series, where $\ell$ is $s$ or $d$. In the case of the $^2D^o$ kernel, four Rydberg
series are possible: $^3D^o$ from the $ns$-Rydberg orbitals and $^3S^o$, $^3P^o$, and $^3D^o$
from the $nd$ Rydberg orbitals.  For the $^2P^o$ core there are three optically allowed series; 
$^3P^o$ from the $ns$-Rydberg orbitals and $^3P^o$, $^3D^o$ from the $nd$-Rydberg orbitals. 
%
%
%
\begin{figure*}
\begin{center}
\includegraphics[scale=1.5,width=\textwidth]{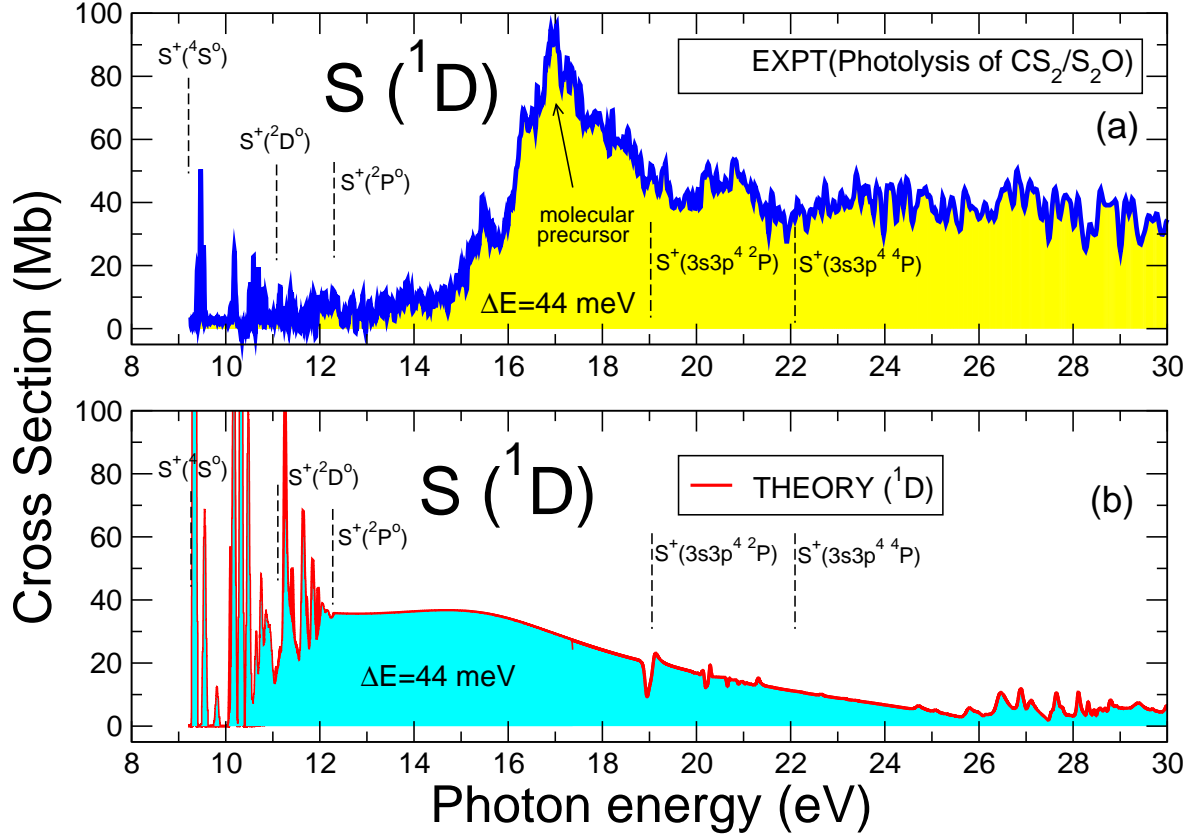}
\caption{\label{1D}  (Color online) An overview of the experimental measurements on S$(^1D)$
as a function of the photon energy; (a) experiment for the S$(3s^23p^4~^1D)$ metastable state
over the energy range from threshold to 30 eV. Primary photolysis of \ce{CS2} was carried out 
at $\lambda$ =193 nm   {(6.424 eV)}.  Contributions from ionic fragmentation 
photolysis of \ce{CS2} are not subtracted. The curve is 
normalized to the VUV-photon flux.  
(b) Theoretical cross section results for    {100\% of}  the S$(3s^23p^4~^1D)$ metastable state
from the DARC calculations convoluted at 44~meV. 
The resonance series limits $E_{\infty}$ of Equation (2) for each series are 
indicated by vertical-dashed lines in the end of the line groups. 
Resonance features from this Figure are tabulated in Table II.} 
\end{center}
\end{figure*}

\subsection{Rydberg resonance series of S$(3s^23p^4~^3P)$}

As illustrated in Figure \ref{thres3P}, the first ionic continuum $^4S^o$ 
manifests itself in the cross section for S$(3s^23p^4~^3P)$, as obtained from the photolysis of \ce{S2O}. There is a sudden
increase in ion yield at a photon energy of 10.35 $\pm$ 0.02 eV, although there is a relatively weak continuum.
This is particularly evident in the decrease in the photoionization cross section value
around 11 eV similarly observed by Gibson and co-workers \cite{Gibson1986}.
The S$^+(^2D^o)$ and S$^+(^2P^o)$ ionic thresholds (respectively at 12.20 eV and 13.40 eV), 
are partially overlaid with autoionizing Rydberg states.
In the present work, detailed measurements were made up to the photon energy of 13.40 eV,
for the case of resonance series converging to the S$^+(^2P^o)$ threshold from the 
initial S$(3s^23p^4~^3P)$ state.
For the S$(3s^23p^4~^3P)$ state, shown in Figure \ref{thres3P},
strong resonance series are observed in the photon energy range 12.20 - 13.4 eV, i.e.
below the $S^+(^2P^o)$ threshold. 
The assignments and positions of these Rydberg resonance 
series are in suitable agreement with the NIST tabulations \cite{nist}.

It is expected that all members of $3s^33p^3(^2P^o)n\ell$ Rydberg resonance 
series (with the exception of the $3s^33p^3(^2P^o)ns~^3P^o$ series), 
converging to the S$^+(^2P^o)$ threshold are observed.
Members of this group were seen in the early work of Gibson and co-workers \cite{Gibson1986} 
as shoulders of the $3s^33p^3nd~^3P^o$ series but are not observed in the present work 
which is primarily due to the limited energy resolution of the tunable VUV plasma source.
Individual members of the various $3s^23p^3(^2D^o)n\ell$ series;
$nd~^3D^o$, $ns~^3D^o$, $nd~^3S^o$ and $nd~^3P^o$ series can be identified. 
At photon energies of 10.557 eV, 10.561 eV, and 10.646 eV
we find the triplet Rydberg states $3s^23p^3 (^2D^o) 3d ~^3P^o$, 
$3s^23p^3 (^2D^o) 5d ~^3P^o$, and $3s^23p^3 (^2D^o) 6d ~^3P^o$. 
However, due to the narrow line width of the resonances and the 
limited energy resolution (44 meV FWHM), many of the 
resonances were not resolved. A similar situation 
occurred for the resonances associated with the
$3s^33p^3(^2P^o)nd~^3P^o$ and 
the $3s^33p^3(^2P^o)nd~^3D^o$ Rydberg resonance series.
%
\begin{figure*}
\begin{center}
\includegraphics[scale=1.5,width=\textwidth]{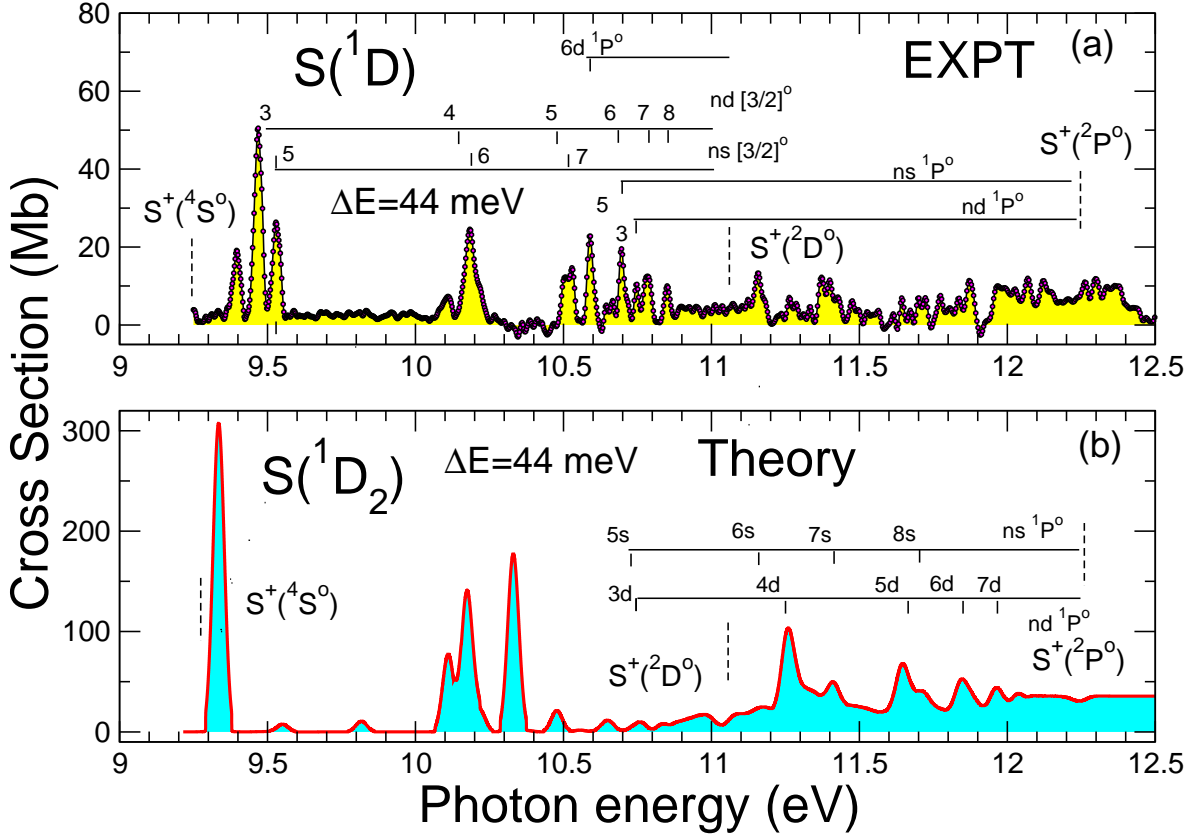}
\caption{\label{thres1D} (Color online) Single photon ionization of sulfur as a
function of the photon energy. 
  {(a)} The experimental cross section is recorded with a nominal 
energy resolution of 44 meV. 
The assigned Rydberg series are indicated by vertical lines grouped by 
horizontal or inclined lines. 
The corresponding series limits $E_{\infty}$ of Equation \ref{eV} for 
each series are indicated by vertical-dashed lines at the end of the line 
groups. 
The first few values of $n$ for each series is displayed close to its 
corresponding vertical 	
line in each group.   {(b)
Single photoionization of atomic sulfur as a
function of the photon energy in the $3s^23p^4~^1D$ metastable state from threshold to 12.5 eV. 
Theoretical cross section were carried out with the DARC codes,
and convoluted with a Gaussian having a profile of 44 meV. 
The assigned Rydberg series are indicated as vertical lines grouped by horizontal or inclined lines. 
The corresponding series limits $E_{\infty}$ of Equation \ref{eV} for 
each series are indicated by a vertical-dashed lines in the end of the line groups. 
Resonance energies and quantum defects for 
the series lying below the S$^+(^2D^o_{3/2, 5/2})$ thresholds
are tabulated in Table \ref{meta-1D}.} }
\end{center}
\end{figure*}

\subsection{Window resonances of S$(3s^23p^4~^3P)$}

In atomic sulfur, the $3s3p^4~^4P$ and $3s3p^4~^2P$ ionization states 
(occurring from a 3s vacancy state) 
are located at 19.056 eV and 22.105 eV, respectively. In the observed 
photoionization spectra 
of atomic sulfur for the $(3s^23p^4~^3P)$ ground state (in the photon energy 
range 17.8 -- 20.2 eV) 
we observe two prominent members of a Rydberg series of window resonances. 
These features were also seen in the early theoretical work of Altun and 
co-workers \cite{Altun1992}.
These dip-like structures were first observed in this photon region by 
Innocenti and co-workers \cite{Innocenti2007} 
using constant ionic state (CIS) spectroscopy. Innocenti and co-workers 
\cite{Innocenti2007} 
showed that these window resonance originate from the $3s^23p^4$ parent 
configuration and are 
due to the $3s \rightarrow np$ excitation being members of the $3s3p^4np$ 
Rydberg resonance series.

The ground state electronic configuration of the sulfur atom is 
$(3s^23p^4~^3P)$, i.e. 
valence shell photoionization from this state produces the ionic continuum 
states S$^+(^4S^o)$, S$^+(^2D^o)$ and S$^+(^2P^o)$, which have
ionization potentials of 10.36 eV, 12.20 eV, and 13.40 eV \cite{nist}. 
Inner-shell $3s$ ionization results in the S$^+$ ionic $3s3p^4~^4P$ and 
$3s3p^4~^{  {2}}P$ hole states that
have ionization potentials of 20.20 eV and 23.45 eV, respectively \cite{nist,Innocenti2007}. 
The window resonances occur due to excitation first into a 
Rydberg resonance state caused by autoionization, 
then into the energetically accessible $^4S^o$, $^2D^o$, and $^2P^o$ continua 
transitions.
The Rydberg states involved are triplet states, so the oscillator strengths 
for transitions
violate the spin-selection rule $\Delta S$ = 0, which in the case of atomic 
sulfur is very low. 
The most intense transitions occur for $\Delta L$ = 0, $\pm$ 1, in this case 
$ns \rightarrow np$ transitions are expected to occur.

  {The dipole selection rule ($\Delta L$ = 0, $\pm$ 1) limit the allowed Rydberg 
resonance states to the
$3s3p^4 (^4P) np~^3D^o$, $3s3p^4 (^4P) np~ ^3P^o$ and $3s3p^4 (^4P) np~ ^3S^o$ 
resonance series,
where the assignments are based on earlier work of Innocenti et al. \cite{Innocenti2007}.
Their experimental 
work 
has already revealed the position of members of these resonance
series.
In the present study the first two members of the resonance series were 
observed at
17.82 $\pm$ 0.09 eV (69.57 $\pm$ 0.35 nm) and 19.02 $\pm$ 0.09 eV (65.19 $\pm$ 
12.35 nm), respectively, which is in good agreement 
with the experimental 
work of Innocenti et al. \cite{Innocenti2007}. }
Our present theoretical estimates are also in respectable agreement with the 
present experimental work.
The resonances are very intense and couple to the $^2D^o$ and $^2P^o$ continuum.
Other members of this Rydberg resonance series are    {obtained} 
at 19.55 $\pm$ 0.09 eV and 19.78 $\pm$ 0.09 eV
which agree well with the results of Innocenti et al. \cite{Innocenti2007}. 
Innocenti and co-workers \cite{Innocenti2007} observed that both of the $n$=3 
and 4 members
of this Rydberg resonance series have a spectral line-width just below the 
experimental resolution.
In the present work, the window resonances are observed 
by detection of the S$^+$ ions from photoionization of atomic sulfur in the S$(3s^23p^4~^3P)$ ground state.

  {Similar types of structures were observed previously 
by Angel and Samson \cite{Samson1988} for atomic oxygen. 
 Window resonances were also observed for atomic selenium, by  
 Gibson et al. \cite{Gibson1986} and atomic tellurium,  by Berkowitz et al. \cite{Berk1981}.} 
However, in all these previous studies, the structures 
are clearly less intense than in the present case of atomic sulfur.

\subsection{Photoionization of metastable S$(3s^23p^4~^1D~ {\rm and}~ ^1S)$} 

The photoion yield of atomic sulfur in the S$(3s^23p^4~^3P)$ state, 
using \ce{CS2} as a precursor, 
allows us to obtain photoion yield curves of atomic sulfur, in the S
$(3s^23p^4~^3P)$ and S$(3s^23p^4~^1D)$ states. 
By subtraction means, the excited metastable state of atomic sulfur S
$(3s^23p^4~^1D)$ is then accessible.
The excited S$(3s^23p^4~^1D)$ metastable state of atomic sulfur is of 
fundamental spectroscopic interest because here the behavior 
of electronically excited states of systems can be analyzed.  
The S$(3s^23p^4~^1D)$ metastable state of atomic sulfur is located at 1.145 eV 
above 
the S$(3s^23p^4~^3P)$ ground electronic state, where the first three 
ionization energies (for the 
$^4S^o$, $^2D^o$ or $^2P^o$ continuum) are located at 9.215 eV, 11.060 eV, and 
12.25  {5} eV, respectively. 

Previous experimental studies on the photoionization of atomic sulfur in the 
first electronically excited metastable
state S$(3s^23p^4~^1D)$ are rather limited in energy range. The work of Pan 
and co-workers \cite{Pan2008} 
studied the energy 75 800 -- 89 500 cm$^{-1}$ (9.4 - 11.1 eV), using 
dissociation of \ce{CS2} after photolysis 
at 193 nm   {(6.424 eV)} produced sulfur atoms in a singlet excited states which were 
subsequently ionized by synchrotron radiation. 
A resolution of up to 3 cm$^{-1}$ (0.37 meV) was attainable at the NSRRC 
beamline U9CGM in China. 
Several Rydberg resonances series were detected and analyzed converging to the 
S$^+(^2D^o_{3/2, 5/2})$ 
thresholds.  

On the theoretical side several studies of atomic sulfur dealing with the 
transitions and energies 
of the ground $3s^23p^4~^3P_2$ and metastable $3s^23p^4~^1D_2$ states
are available in the literature as mentioned in the Introduction.
Experiments were first performed by Joshi and co-workers \cite{Joshi1987} 
for the case of the sulfur atom in the $3s^23p^4~^1D_2$ state for the energy 
range
below the S$^+(^2P^o)$ threshold, which is located at 12.255 eV. Two Rydberg series 
were detected in the gas phase absorption spectrum. 
The photon energy range studied in previous experiments carried out
below the S$^+(^2D^o)$ threshold (located at 11.060 eV) is rather limited. 
The work of Qi et al. \cite{Qi2002} investigated photoionization 
of atomic sulfur in the initial S$(3s^23p^4\;^1D)$ state,
from the 193 nm   {(6.424 eV)} excitation of thiethane (\ce{C3H6S}) 
up to a photon energy of 13.5 eV using synchrotron radiation 
at 200~meV FWHM resolution. However, the energies of their observed Rydberg resonances series 
 neither agree with the findings of other authors nor with the present work.
%

\begin{figure*}
\begin{center}
\includegraphics[scale=1.5,width=\textwidth]{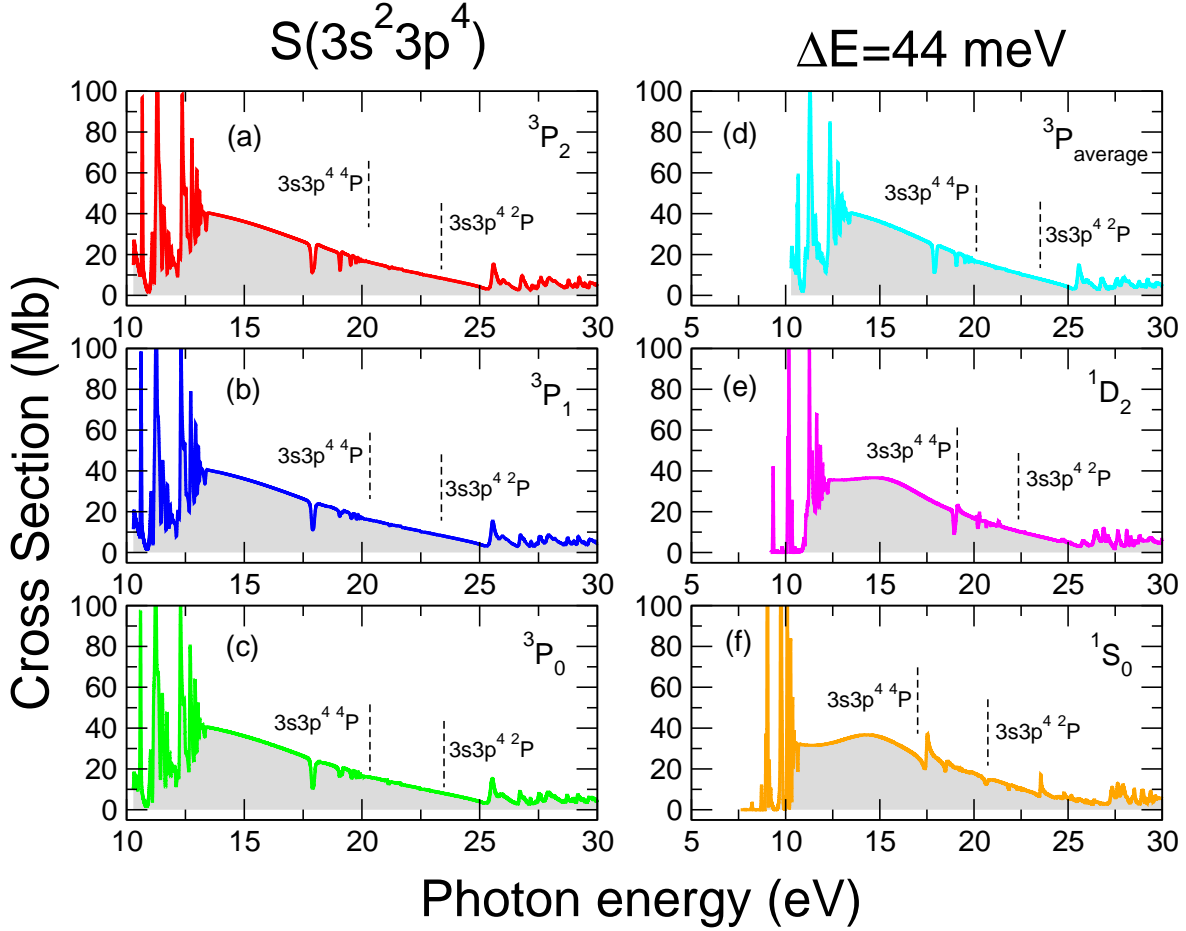}
\caption{\label{all} (Color online) 
Theoretical cross sections from the 512 level DARC calculations for sulfur $3s^23p^4~ ^3P_{2,1,0}$, 
and $3s^23p^4~ ^1D_{2}$, and $3s^23p^4~ ^1S_{0}$ initial states convoluted with a Gaussian profile of 44 meV. 
Single photoionization cross sections of the sulfur atom as a function of energy over the photon energy from thresholds to 30 eV 
illustrating strong resonance features in the spectra below 14 eV.
(a) $3s^23p^4~ ^3P_2$, (b) $3s^23p^4~ ^3P_1$ (c) $3s^23p^4~ ^3P_0$, (d) $3s^23p^4~ ^3P$ 
level averaged, (e) metastable $3s^23p^4~ ^1D_2$
and (f) metastable $3s^23p^4~ ^1S_0$ cross sections.
The corresponding series limits $E_{\infty}$ of Equation \ref{eV} for the window resonances converging
to $3s3p^4~ ^4P$ and $3s3p^4~ ^2P$ sulfur ion thresholds
are indicated by vertical-dashed lines.}
\end{center}
\end{figure*}

In the photon energy region below 11.060 eV, the use of the $^2D^o$ continuum
relative to the energy of S$(3s^23p^4~^1D)$, has been studied 
by a variety of different authors where the known
autoionizing Rydberg resonances series were resolved.
Experiments were primarily carried out by tunable VUV radiation
employing frequency tripling of UV laser radiation. The high resolving power permitted to determine
 the lifetimes of some
of the excited Rydberg states. A disadvantage of this approach though
is the very limited frequency (energy) range investigated, as it excludes photoionization studies 
at higher photon energies and broad tuning of the radiation.

%
%
%
%
\begin{table*}
\caption{\label{meta-1D} Principal quantum numbers $n$, resonance energies (eV), and 
					quantum defects $\mu$ of the prominent S$(3s^23p^3 [^2D^{\circ}_{5/2,3/2}] ) ns,nd$ Rydberg series seen in the 
					S$(3s^23p^4~^1D_2)$ photoionization spectra converging to the S$^+(3s^23p^3 [^2D^{\circ}_{3/2,5/2}])$ and
					S$^+(3s^23p^3 [^2P^{\circ}])$ thresholds.
					The experimental resonance energies are calibrated to $\pm$20 meV and 
					quantum defects $\mu$ are estimated to within an error of 20\%.
					The assignments are shown in Fig. \ref{thres1D}. The theoretical results are obtained from the 512-level DARC
					calculations performed within the Dirac Coulomb {\it R}-matrix approximation. The experimental spectral assignments, 
					from the thesis work of Barthel \cite{barthel2009}, are uncertain for entries in parentheses.} 
\begin{ruledtabular}
\begin{tabular}{cccccccccc}                                         
Sulfur		&			&$E_{n}$ (eV) 				&$E_{n}$ (eV)			&$\mu$			& $\mu$		& $E_n$ (eV) 			& $E_n$ (eV)	&$\mu$		  & $\mu$ \\
(Initial state) 	& 		  	& (Expt) 					&(Theory)				&(Expt)			&(Theory)  	&(Expt)	  			& (Theory)	&(Expt)		  &(Theory)\\  
\hline                                      							       \\
& $ns$  		&   &$3s^23p^3(^2D^{\circ}_{3/2})ns$	& [$3/2$]  				&	& 		& $3s^23p^3(^2D^{\circ}_{5/2})ns$		& [$5/2$] 	&    \\  
$3s^23p^4~^1D_{2}$ &    	& 		   				&					&				&        	 	& 					&			&				&    \\
& 5    		& (\;9.5300)$^{a}$     		&\;9.5515				&2.02 $\pm$ 0.40	&1.99  		&--        			& --			& --				&    \\
& 6    		& (10.2046)$^{a}$      		&10.2399				&2.01 $\pm$ 0.40	&1.93  		&--        			&10.2192		&				&0.98\\
& 7    		& (10.5136)$^{a}$     		&10.5125				&2.00 $\pm$ 0.40	&2.01  		&--        			&10.5115		&				&2.02\\
& 8    		& --      					&10.6848				&--				&1.98	  	&--        			&10.6785	 	& --				&2.03\\
& 9    		& --      					&10.7840				&--				&1.98	  	&--        			&10.7793	 	& --				&2.04\\
& 10    		& --      					&10.8485				&--				&1.98	  	&--        			&10.8442	 	& --				&2.07\\
& 11    		& --      					&10.8929				&--				&1.97	  	&--        			&10.8891	 	& --				&2.08\\
& $\cdots$ 	& $\cdots$    				&$\cdots$				&--				& $\cdots$ 	&$\cdots$        		&$\cdots$		&$\cdots$			&$\cdots$\\
& $\infty$ 		& 11.0599$^{b}$    			&11.0599$^{b}$ 	&--				& 		    	&11.0603$^{b}$	&11.0603$^{b}$&			&	\\ 
\\
\\
& $nd$     	&  &$3s^23p^3(^2D^{\circ}_{3/2})nd$   		&[$3/2$]			&     & 		&$3s^23p^3(^2D^{\circ}_{5/2})nd$	&[$5/2$]		&     \\  
$3s^23p^4~^1D_{2}$ &    	& 	   					&					&				&        		& 					&			&				&     \\
& 3    		& (\;9.4690)$^{a}$     		&\;9.4722				&0.08 $\pm$ 0.02	&0.07	  	& --       			&--			&				&     \\
& 4    		& (10.1715)$^{a}$     		&10.1739				&0.09 $\pm$ 0.02	&0.08  		& (10.08762)$^{a}$&10.1112		&0.26 $\pm$ 0.05 	&0.21 \\
& 5    		& (10.4926)$^{a}$    			&10.4778				&0.10 $\pm$ 0.02	&0.16  		& --        			&10.4759		&--				&0.18 \\
& 6    		& (10.6706)$^{a}$    			&10.6575				&0.09 $\pm$ 0.02	&0.18  		&--        			&10.6548		&--				&0.21 \\
& 7    		& (10.7736)$^{a}$      		&10.7674				&0.11 $\pm$ 0.02	&0.18	  	&--        			&10.7634		&--				&0.23 \\
& 8    		& (10.8406)$^{a}$      		&10.8338				&0.12 $\pm$ 0.03	&0.24	  	&--        			&10.8363		&--				&0.21 \\
&  9   		&  --     					&10.8842				& --				&0.20	  	&--        			&10.8823		&--				&0.26 \\
& 10    		&  --     					&10.9179				& --				&0.21	  	&--        			&10.9161		&--				&0.28 \\
&  11		& --							&10.9430				& --				&0.21		&--				&10.9415		&--				&0.29\\
& $\cdots$ 	& $\cdots$    				&$\cdots$				& --				& $\cdots$ 	&$\cdots$        		&$\cdots$			&$\cdots$			&$\cdots$\\
& $\infty$ 		& 11.0599    				&11.0599	$^{b}$ 		&				& 		    	&11.0603	$^{b}$	&11.0603	$^{b}$	&		&	\\ 
\\
\\
& $n$     	&  &$3s^23p^3(^2P^{\circ})ns$   		&[$^1P^o$]	&     			& 		&$3s^23p^3(^2P^{\circ})nd$			&[  {$^1P^o$}]		&     \\  
$3s^23p^4~^1D_{2}$ &    	& 	   					&					&				&        		& 					&			&				&\\
&3			&						&					&				&			&10.741359$^{c}$		&10.7397		&0.007			&0.008 \\
&			&						&					&				&			&10.741393$^{d}$		&--			&0.007			&--	\\
&4			&--						&					&--				&			&--					&11.2587		&--				&0.30 \\
&5			&10.712359$^{a}$			&10.71400			&2.03			&2.03		&--					&11.6536		&--				&0.26\\
&			&10.712257$^{b}$			&--					&2.03			&--			&--					&--			&--				&--	\\
&6			&--						&11.1675 				&--				&2.02		&--					&11.8519		&--				&0.22\\
&7			&--						&11.4017				&--				&2.06		&--					&11.9688		&--				&0.16\\
&8			&						&11.7017				&--				&--			&--					&--			&--				&--	 \\
&$\cdots$		&						&$\cdots$				&--				&--			&$\cdots$				&--			&--				&$\cdots$	\\
& $\infty$     	& 					   	&12.2598$^{b}$  		&--				&--		  	&12.2598$^{b}$ 		&12.2598$^{b}$&			&     \\
\end{tabular}
\end{ruledtabular}
\begin{flushleft}
$^{a}$Experimental work from the thesis of Barthel \cite{barthel2009}.\\
$^{b}$Rydberg series limits $E_{\infty}$ for the sulfur ion (S$^+$) are from the NIST tabulations \cite{nist}.\\
$^{c}$NIST tabulations \cite{nist}.\\
$^{d}$Experimental work of Pan and co-workers \cite{Pan2008}.\\
\end{flushleft}
\end{table*}

In the present work, by appropriate subtraction of the photo-ion yield curve,
from the photolysis process of \ce{CS2}, both states of the sulfur species, 
S$(3s^23p^4~^3P)$ and S$(3s^23p^4~^1D)$ can be studied.
The photo-ion yield curve of atomic singlet sulfur S$(3s^23p^4~^1D)$
was determined in the photon energy range up to 30 eV.
The analysis of the relative intensities of the autoionization resonances,
originating from S$(3s^23p^4~^1D)$ and S$(3s^23p^4~^3P)$ 
states that are formed by photolysis of \ce{CS2},
aids to determine both sulfur states ($3s^23p^4~^3P$ and $3s^23p^4~^1D$). 
  {
From our experimental studies, Rydberg resonances associated with the S$(3s^23p^4~^1D)$ 
initial atomic state are found to be located at 9.47 $\pm$ 0.02 eV and at 10.21 $\pm$ 0.02 eV, respectively.
Similarly for the $S(3s^23p^4~^3P)$ state, resonances are found located at 10.72 $\pm$ 0.02 eV,
11.34 $\pm$ 0.07 eV and 11.66 $\pm$ 0.07 eV.
The relatively high margin of error arises from the uncertainty in
the scaling of the absolute photoionization cross-section to the theoretically
calculated values. }

  {
The branching ratio of these atomic sulfur species 
formed by photolysis can be reliably determined. 
Under the present experimental conditions we find 
the branching ratio ${\rm S}(^1D_J)/{\rm S}(^3P_J)$ = 2.8 $\pm$ 0.4 
at a photolysis wavelength of 193 nm   {(6.424 eV)}.
}
The ratio of 2.8 $\pm$ 0.4 found in favor of
of S$(3s^23p^4~^1D)$ is in clear contradiction to earlier work, where S$(3s^23p^4~^3P)$ 
is the more abundant species.
Waller et al. reported a branching ratio ${\rm S}(^3P_J)/{\rm S} (^1S_J)$ = 2.8 $\pm$ 0.3 \cite{Waller1987} 
and 1.6 $\pm$ 0.3 was reported by Xu et al. \cite{Jackson2004}, also in favor of S$(^3P)$.
However, the present value of 2.8 $\pm$ 0.4 in favor of the S$(3s^23p^4~^1D)$ state
agrees with that found by Yang et al \cite{Jackson2008a,Jackson2008b}.
There is no plausible explanation for the large discrepancy
in the experimental values for the branching ratio of S$(3s^23p^4~^1D)$ to
S$(3s^23p^4~^3P)$ from the 193 nm   {(6.424 eV)} photolysis of \ce{CS2}. It is assumed
that likely different pulse energies of the laser photolysis produce different
product distributions. 

Note, a high photon density would lead to multi-photon effects, with a branching ratio
that may differ significantly from that of one-photon excitation conditions. 
Pulse energies of 1-2 $mJ/cm^2$ \cite{Jackson2004} may cause
additional contributions of S$(3s^23p^4~^3P)$ by the dissociation of vibrationally excited \ce{CS}. 
The pulse energy of the 193 nm   {(6.424 eV)} excitation radiation, used in the present work, 
is about 0.4 ${\rm mJ/cm^2}$, so we safely assume that multi-photon effects play a minor role.

In Figure \ref{thres1D} the energies of the $^4S^o$, $^2D^o$ and $^2P^o$
thresholds for the \ce{S^+} ion relative to the S$(3s^23p^4~^1D)$ threshold 
are located at 9.215 eV ($^4S^o$), 11.060 eV ($^2D^o$), and 12.256 eV ($^2P^o$), respectively.
These thresholds are indicated in Figure \ref{thres1D} by vertical dashed lines. It should be noted that direct
photoionization is spin-forbidden in the $^4S^o$ - continuum because of the
totality of the $^4S^o$-continuum and the emitted photoelectron. 

In the energy range below 12.5 eV, autoionizing Rydberg series are observed,
converging to the energetically higher lying ionic continua.
These two allowed ionic continua S$^+ (^2D^o)$ and
S$^+ (^2P^o)$ are located at 11.060 eV and 12.256 eV, respectively. 
A significant increase in photoionization efficiency is recorded 
above a photon energy of about 14.6 eV (c.f. Figure \ref{1D} (a)). 
There is a wide maximum structure stretching to about 17 eV with a shoulder at about 15.4 eV. 
This would appear to be due to residual molecular effects from the photolysis process, as outlined in 
Table \ref{tabAppErg} (cf. Section IV B). Evidence for this assumption comes from the fact that 
these features do not show up in the corresponding theoretical work on atomic sulfur from the excited 
$3s^23p^4~^1D_2$ metastable state as illustrated in Figure \ref{1D} (b).
In the energy region below the $S^+ (^2P^o)$ threshold 
the assignment of the resonance features in the photoionization spectra 
are in good agreement with those available in the literature. The respective Rydberg states
couple via the process of autoionization to the ionic $^4S^o$ continuum.
Between the individual resonances, the $S^+$ yield falls almost to zero.

The present study gives resonant structures assignment                 
for the Rydberg series converging to the S$^+(^2D^o_{3/2, 5/2})$ ionic thresholds.
Autoionizing resonance states are observed for the main Rydberg series, converging to 
the S$^+(^2D^o)$ ionic continuum shown in Figure \ref{thres1D} (a) and (b). The most intense of
the resonance series is the $3s^23p^3 (^2D^o) nd~ ^1D^o$. Here, the transitions can
be resolved for the $n$ = 3, 4, 5, 6, 7, and 8 members of the Rydberg resonance series 
(see Table \ref{meta-1D}). 
The higher members are not observed due to the low intensity and the limited experimental 
resolution of the VUV radiation. Theoretical predictions from the 512-level DARC 
calculations (given in Table \ref{meta-1D}) are seen to be in good 
agreement with the experimental measurements and with the 
measurements of Pan and co-workers \cite{Pan2008} (i.e., Series II).
All the autoionizing Rydberg transitions occurring in atomic sulfur from the $3s^23p^4~^1D_2$ 
excited metastable state, at energies below the $^2D^o_{3/2, 5/2}$ thresholds, are seen
to be in good agreement with the values available from the literature \cite{nist,Pan2008}. 

Rydberg series members are observed for the $3s^23p^3 (^2D^o) ns~ ^1D^o$ 
resonances series with principal quantum numbers $n$ = 5, 6, and 7. This series 
was also observed by Pan and co-workers \cite{Pan2008} (i.e. Series I).
We note that the $n$ = 6 and $n$ = 7 Rydberg members partially overlap.
The resonance found at 10.59 $\pm$ 0.02 eV is from the 
$3s^2 3p^3 (^2D^o) nd~ ^1P^o$ Rydberg series where $n$ = 6. 
This transition appears unusually intense and there 
are no other members of this group observed. The unusually
high intensity of this transition is due to the energetic proximity
explained by the triplet Rydberg states and the interference with the singlet state. 

In the present work we observe for the first time
autoionizing Rydberg resonance states above the $^2D^o$ threshold
and lying below the $^2P^o$ continuum. These resonances are also seen in the 
DARC photoionization cross section calculations for the ground 
and metastable states (see Figure \ref{all}).
The excited Rydberg resonance states in the cases discussed so far are all
$^1D^o$ and $^1P^o$ states. The ionic continuum to which these states
couple in autoionization to is a $^4S^o$ continuum. The
spin $S_J$ of the ionic core is 3/2. Secondly, if one considers the
totality of the ionic core and the emitted electron with spin $s$ =$\pm$1/2, 
a total spin $S$ of 1 or 2 is possible.
This will give rise to triplet and quintet-terms, while the 
autoionizing Rydberg states form singlet terms. 
Taking into account the selection rules for autoionization
$\Delta S=0$ appears initially banned for this transition. Under
spin-orbit coupling we can expand the selection rule of the autoionization
however, to $\Delta S$ = 0, $\pm$1, $\pm$2. For transitions of this type,
due to the spin-orbit coupling, large resonances appear
having relatively narrow width.
%

\begin{figure*}
\begin{center}
\includegraphics[scale=1.5,width=\textwidth]{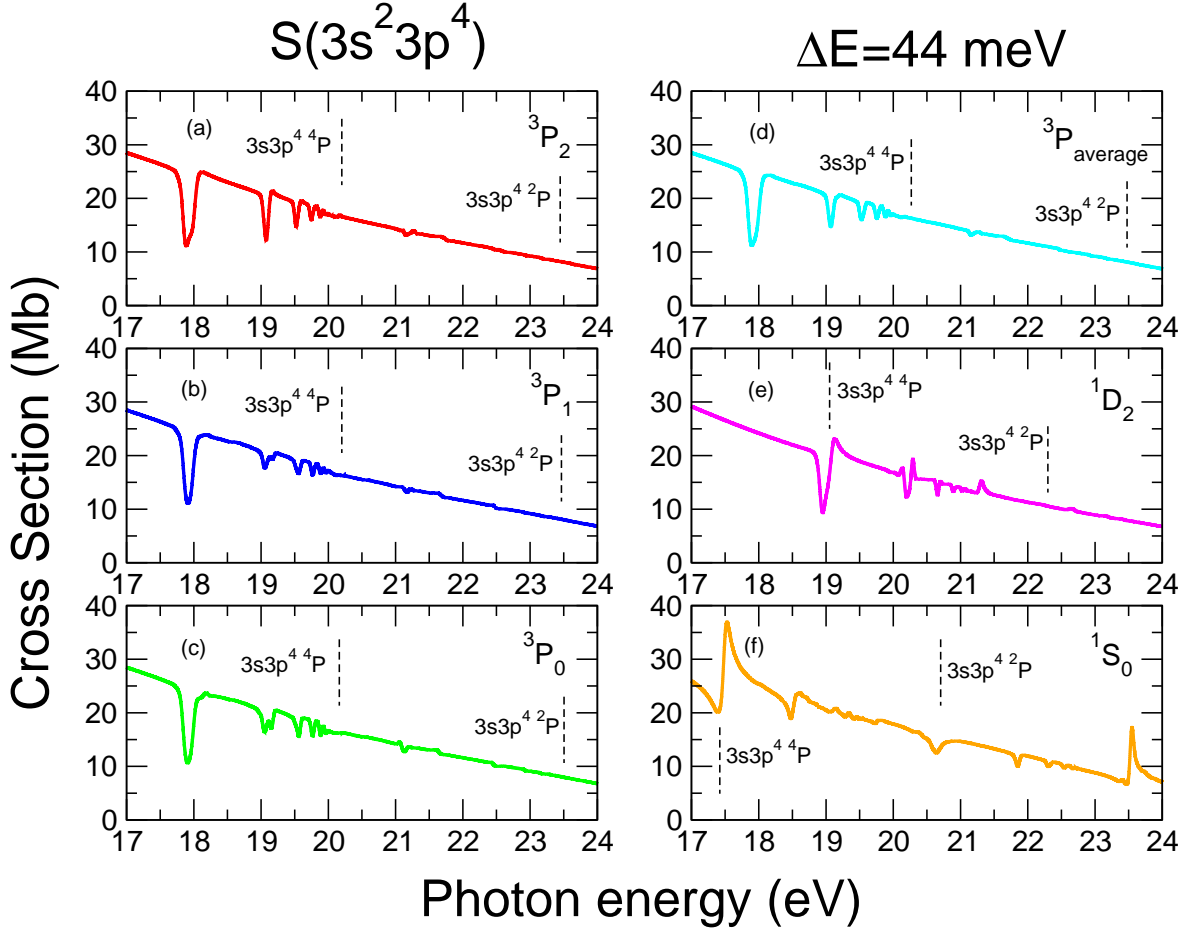}
\caption{\label{window} (Color online) Theoretical cross sections results 
from the 512 level DARC for the sulfur $3s^23p^4~ ^3P_{2,1,0}$, 
and $3s^23p^4~ ^1D_{2}$, and $3s^23p^4~ ^1S_{0}$ initial states, 
in the photon energy region 17 -- 24 eV, 
convoluted with a Gaussian having a profile of 44 meV. 
(a) $3s^23p^4~ ^3P_2$, (b) $3s^23p^4~ ^3P_1$ (c) $3s^23p^4~ ^3P_0$, 
(d) $3s^23p^4~ ^3P$ level averaged, (e) metastable $3s^23p^4~ ^1D_2$,
and (f) metastable $3s^23p^4~ ^1S_0$ cross sections.
In this energy region the prominent window resonances converging to the singly ionized sulfur ion threshold 
$3s3p^4~ ^4P$ (lowest vertical dashed line) are clearly visible in the cross sections. See text for further details.}
\end{center}
\end{figure*}

The autoionization resonance occurs with a maximum at a photon energy of
10.11 $\pm$ 0.02 eV. However, it appears broadened 
and can be assigned to the $3s^23p^3(^2D^o) nd~ ^3D^o$ series. 
Here, for the process of autoionization, the selection rule 
$\Delta S=0$ follows, as no spin-orbit coupling effects are involved. 
The lifetime will therefore be significantly reduced and the autoionization resonances
appear much broadened. It should be noted that the
excitation of the $3s^23p^4~^1D$ state in a $3s^23p^4~^3D$ Rydberg state to the 
spin-selection rule in the context of one-photon processes is contrary, illustrated by
the relatively low intensity. This transition was also observed 
by Yang and co-workers \cite{Jackson2008a,Jackson2008b} 
who indicated that spin-spin and spin-orbit interactions in the case of atomic sulfur may play a role.

%
%
%
%
\begin{table*}
\caption{\label{meta-1S} Principal quantum numbers $n$, resonance energies (eV), and 
					quantum defects $\mu$ of the prominent S$(3s^23p^3 [^2P^{\circ}] ) ns,nd$ Rydberg series seen in the 
					S$(3s^23p^4~^1S_0)$ photoionization spectra converging to the S$^+(3s^23p^3 [^2P^{\circ}])$ thresholds.
					The assignments are shown in Figure \ref{thres1S-theory}. 
					The theoretical results were obtained from the 512-level DARC
					calculations performed within the Dirac Coulomb {\it R}-matrix approximation.} 
\begin{ruledtabular}
\begin{tabular}{cccccccccc}                                         
Sulfur		&			&$E_{n}$ (eV) 				&$E_{n}$ (eV)			&$\mu$			& $\mu$		& $E_n$ (eV) 			& $E_n$ (eV)	&$\mu$		  & $\mu$ \\
(Initial state) 	& 		  	& (Expt) 					&(Theory)				&(Expt)			&(Theory)  	&(Expt)	  			& (Theory)	&(Expt)		  &(Theory)\\  
\hline                                      							       \\
\\
& $n$     	&  &$3s^23p^3(^2P^{\circ})ns$   		&[$^1P^o$]	&     & 			&$3s^23p^3(^2P^{\circ})nd$	&[$^1P^o$]		&     \\  
$3s^23p^4~^1S_{0}$ &    					& 	   			&	&			&        			& 				&			&				&     \\
& 3			&--						&--					&--				&--			&\;9.1366$^{a}$	&\;9.1361		&0.004$^{a}$	&0.005\\
& 4    		& --    					&--					&--				&--	  		&--				&--			&--			 	&-- \\
& 5    		&\;9.1075$^{a}$    			&9.02574				&2.033$^{a}$ 		&2.109  		& --        			&--			&--				&-- \\
& 6    		& --					    	&9.75287				&--				&2.114	  	&--        			&			&--				&-- \\
& 7    		& --				     		&10.0852				&--				&2.108		&--        			&			&--				&-- \\
& 8    		& --					     	&10.2592				&--				&2.127		&--        			&			&--				&-- \\
& 9			&--						&10.3743				&--				&2.021		&--				&			&				&--\\
& 10			&--						&--					&--				&--			&--				&			&				&--\\
& $\cdots$ 	& $\cdots$    				&$\cdots$				& --				& $\cdots$ 	&$\cdots$        		&$\cdots$		&$\cdots$			&$\cdots$\\
& $\infty$ 		& 10.65363    				&10.65363$^{b}$ 		&				& 		    	&10.65363$^{b}$	&10.65363$^{b}$&		&	\\ 
\end{tabular}
\end{ruledtabular}
\begin{flushleft}
$^{a}$Experimental work of Yang et al. \cite{Jackson2008a}.\\
$^{b}$Rydberg series limits $E_{\infty}$ for the sulfur ion (S$^+$) are from the NIST tabulations \cite{nist}.\\
\end{flushleft}
\end{table*}

The autoionization resonance,  located at a photon energy of 9.40 $\pm$ 0.02 eV,
may be assigned to    {a}  known atomic transition of singlet sulfur. 
This structure was also observed by Pan and co-workers \cite{Pan2008} at a wavenumber
of 75 821 $cm^{-1}$ (corresponding to 9.401 eV) and by Yang et al. \cite{Jackson2008a,Jackson2008b} 
for a wavenumber of 75 818 $cm^{-1}$ (corresponding to 9.400 eV). Pan
et al. \cite{Pan2008} discussed an assignment to a $^1G$ series, but because of
$\Delta J = 2$ this was excluded. Yang and co-workers \cite{Jackson2008a,Jackson2008b} suggested the transfer of excitation
to a $3s^23p^3 (^2D^o) 3d$ state, without the corresponding term symbol. 
  {The energy of the photon \ce{ArF} laser
  {(6.42~eV)} is insufficient to 
form atomic sulfur in the excited state S$(^1S)$, since the photodissociation
energy of \ce{CS2} is 397~kJ/mol = 4.114~eV \cite{Darwent1970}   and the excitation energy of
S($^{1}$S) is 2.7499637~eV \cite{nist} so that this electronic state is
excluded as the cause of the resonance.} The resonance lies energetically above the
$^4S^o$ continuum of S$(^3P)$, so the resonance in the photoionization
of the ground-state sulfur is not observed. Against this background and
given the relatively low excitation energy of 9.40 $\pm$ 0.02 eV, tentatively
assigning this feature to a $3s^23p^3 (^2D^o)3d$ state appears straightforward.

Rydberg series converging to the $^2P^o$ continuum are observed 
to be members of the $3s^23p^3 (^2P^o) nd ~^1P^o$ 
and the $3s^23p^3 (^2P^o) ns ~^1P^o$ resonance series.
The autoionizing $3s^23p^3 (^2P^o) 5s ^1P^o$ resonance state occurs at an energy 
of 10.70 $\pm$ 0.02 eV, the $3s^23p^3 (^2P^o) 3d~ ^1P^o$ 
state at a photon energy of 10.75 $\pm$ 0.02 eV. 
These energies are in good agreement with the values from the 
NIST tabulations \cite{nist} of 10.7123 eV and 10.7414 eV.
We note that both resonances were observed by Pan and co-workers \cite{Pan2008}. 
The quantum defect $\mu$ of the $5s$ resonance at an energy 
of 10.70 $\pm$ 0.02 eV is 2.03, while that for the $3d$ resonance 
(at an energy of 10.75 $\pm$ 0.02 eV), is approximately zero. 
Extrapolating with the Rydberg formula using   {(see equation \ref{eV})} and converting all quantities to eV
one   {could straight forwardly} locate  further members of    {the} Rydberg resonance series. 
%
\begin{figure*}
\begin{center}
\includegraphics[scale=1.5,width=\textwidth]{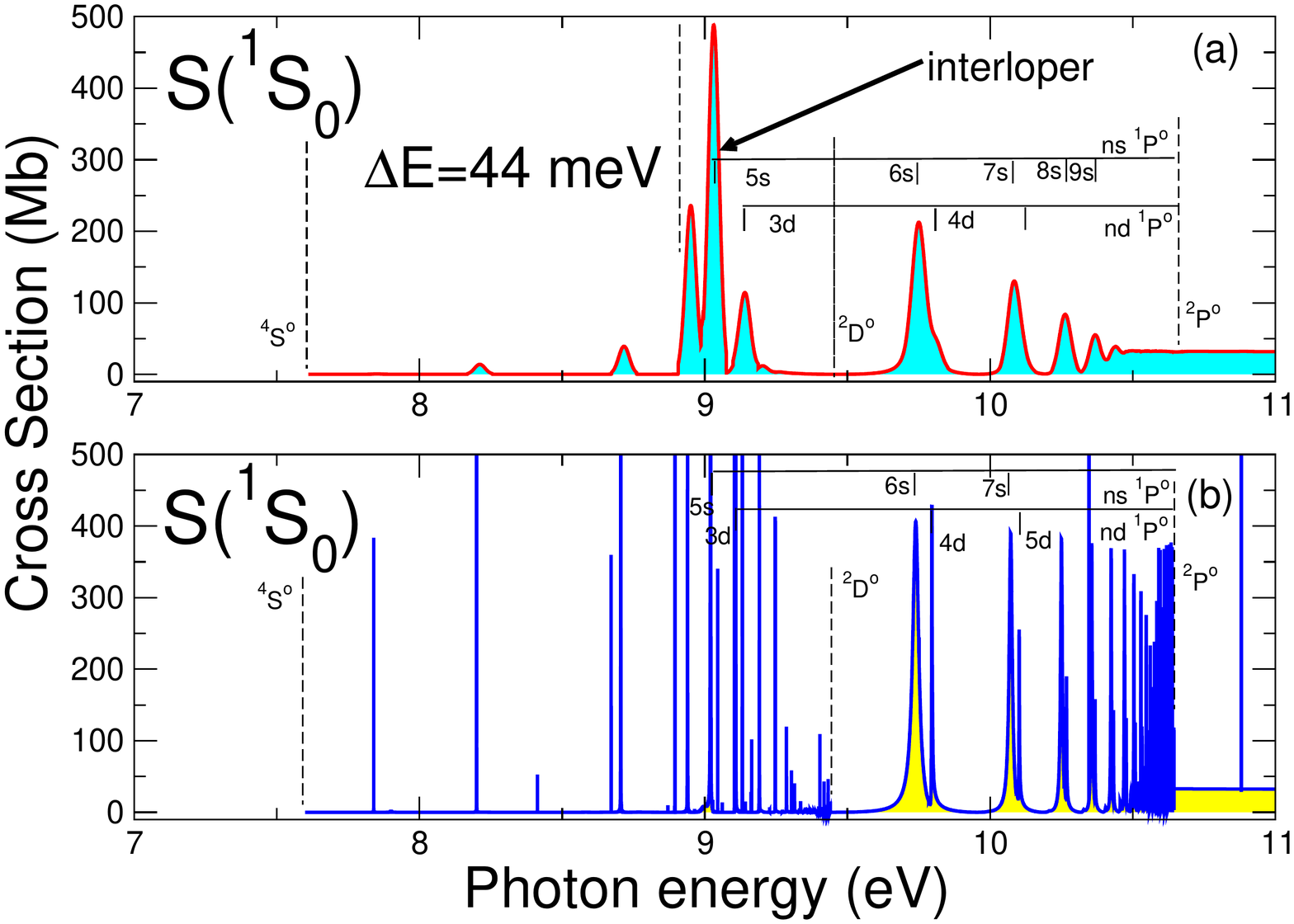}
\caption{\label{thres1S-theory} (Color online) 
Single photoionization of atomic sulfur as a
function of the photon energy in the $3s^23p^4~^1D$ metastable state from threshold to 12.5 eV. 
Theoretical cross section were carried out with the DARC codes,
and convoluted with a Gaussian having a profile of 44 meV. 
The assigned Rydberg series are indicated as vertical lines grouped by horizontal or inclined lines. 
The corresponding series limits $E_{\infty}$ of Equation \ref{eV} for 
each series are indicated by a vertical-dashed lines in the end of the line groups. 
The first few values of $n$ for each series is displayed close to its corresponding vertical 	
line in each group. Resonance energies and quantum defects for 
the various series are tabulated in Table \ref{meta-1D}.}
\end{center}
\end{figure*}

For the energy of the higher lying members of the 
$3s^23p^3 (^2P^o) ns~^1P^o$ and the $3s^23p^3 (^2P^o) nd ~^1P^o$ resonance series,
experimental values of approximately, 11.393 eV for the $3s^23p^3 (^2P^o) 6s~^1P^o$ state and 
11.406 eV for the $3s^23p^3 (^2P^o)4d~^1P^o$ state are obtained. 
These values agree well with the energies of the
structures observed at 11.37 $\pm$ 0.02 eV and 11.40 $\pm$ 0.02 eV. For this
reason we assign these resonances to the $3s^23p^3 (^2P^o) 6s ~^ 1P^o$ and the 
$3s^23p^3 (^2P^o) 4d ~^1P^o$ states. 
The resonance state located at an photon energy of 11.16 $\pm$ 0.02 eV 
is not available from the NIST tabulations \cite{nist} since 
in the energy range above the $^2D^o$ - threshold
no singlet resonance states are listed.  
The photon energy of 11.291 eV is the closest lying triplet state, namely, 
$3s^23p^3 (^2P^o) 4d ~^3P^o$. However, 
  {as illustrated in Figure \ref{thres1D} (b)}, 
the DARC cross section calculations in the region above the S$^+(^2D^o)$ - threshold 
and below the S$^+(^2P^o)$ - threshold, intense singlet resonance states are seen including
shoulder resonances located at approximately similar positions (see Table \ref{meta-1D}).

To the best of the authors knowledge the only experimental study from the S$(3s^23p^4~^1S)$ metastable 
state is that of the work of Yang et al. \cite{Jackson2008b} where the S$(^1D)$
and S$(^1S)$ atoms are produced by 193 nm   {(6.424 eV)} photodissociation of \ce{CS2}.
  { As
mentioned above, the photon energy of a 193~nm (6.424 eV) laser is insufficient to produce
S($^{1}$S) from \ce{CS2} photodissociation. The formation of atomic S in its
$^{1}$S state is likely connected with two-photon absorption processes.}
To complete our study on this system we performed large-scale 
DARC calculations on this metastable state for the photon energy range from threshold (7.61 eV) up to 30 eV. 
Figure \ref{all} (f) shows the cross section for this metastable state as a function photon energy. 
Similar to the work presented earlier we have 
convoluted the theoretical cross sections with a Gaussian having a profile of 44 meV. Here again in the energy region below
12 eV   {(see Figures \ref{thres1S-theory} (a) and (b))}, 
strong resonance features are found in the cross section which will be discussed in the following Sections.

\subsection{Resonances: S$(3s^23p^4~^1D~ {\rm and} ~^1S)$}

The electronic configuration $ns^2 np^4$ belongs to the atomic species
O~$(n=2)$ and S~$(n = 3)$ which we use as the basis to interpret the present results.
Due to the different principal quantum numbers these states will have different
energies. However, in general the energies of the sulfur atom appear 
lower by several eV. In the following, the observed
photoionization efficiency of the atomic structure in the singlet
sulfur S$(3s^23p^4~^1D)$ are interpreted in the region around 17 eV.

If one compares the photoion yield of atomic sulfur with that
of atomic oxygen, it is remarkable that in both cases, the respective
electronic ground state S$(3s^23p^4~^3P)$ and O$ (2s^22p^4~^3P)$ series of the autoionizing
Rydberg transitions are observed converging to the respective $^4P$-threshold
at 20.20 eV (S$^+$) and 28.48 eV ($O^+$) relative to the electronic ground state.
For the two electronically excited species $S(3s^23p^4~^1D)$ and O$ (2s^22p^4~^1D)$ 
a broad autoionization resonance occurs at photon energies 
below the respective energetic position of the $^4P$-continuum.
In the case of sulfur S$(3s^23p^4~^1D)$ metastable state, it has a shoulder on the low energy side. 
The resonance in the case of the metastable atomic oxygen O$ (2s^22p^4~^1D)$ state is a Coster-Kronig
process and may be assigned to the $2s^22p^4 ~^1D \rightarrow 2s2p^5 ~^1P^o$ 
transition with subsequent autoionization. 
The analog $3s3p^5~ ^1P^o$ state of atomic sulfur occurs energetically
at significantly lower energy of 8.95 eV relative to S$(3s^23p^4~^1D)$ state. 
This value is below the ionization energy and represents the reason why autoionization
of this state is not observed. 

The window resonance may be depicted by a Fano profile, leading to
a profile index $q$ of 0.21 $\pm$ 0.01, a half-width $\Gamma$ of 0.60 $\pm$ 0.005 eV
located at an energy of 17.44 $\pm$ 0.006 eV.
The observed intense structure in the experimental data 
is found at a photon energy of 16.92 $\pm$ 0.1 eV (c.f. Figure \ref{1D} (a)). 
An interpretation of this intense structure seen at 16.92 $\pm$ 0.10 eV may be given.
We have already discussed that the $3s3p^5~ ^1P^o$ state is due to $3s$-valence 
excitation of atomic sulfur at higher excitation energies, 
an expected analog excitation of a $3s$-electron to the $np$ orbitals with $n> 3$ can occur.
The electron configuration in the case of inner-shell excitation 
$3s3p^4 np$ forms a multiple of terms. 
Atomic sulfur has the terms $^3P$, $^1D$, $^1S$ resulting from the $3s^23p^4$ configuration.
Term energies for the $3s$ excited states with the exception of the 
$3s3p^4 np~ ^5P^o$ and $3s3p^4 nd~^5D$ Rydberg series are unpublished,
but calculations are available from the OPACITY project \cite{OPACITY}.
The $^5P$-states can be considered as an indication
that the appropriate $3s$ excited singlet terms with similar energies are expected.
An intense autoionization resonance is observed at a photon energy of 16.92 $\pm$ 0.10 eV 
Although it may occur in the autoionizing Rydberg resonances below the $^2D^o$-continuum
a resonance is observed, corresponding to a spin-forbidden 
$3s^23p^4~^1D$ transition, and assigned to a $^3D$ Rydberg state. 
However this resonance relative to the spin-allowed singlet-singlet transitions is significantly less intense. 

A broad resonance is observed (c.f. Figure \ref{1D} (a)) with the peak at a photon energy of
16.92 $\pm$ 0.10 eV and the shoulder at 15.45 $\pm$ 0.10 eV, due to 
$3s^23p^4~ ^1D \rightarrow 3s3p^4np$ excitation associated with subsequent autoionization.
The averaged half-widths are 0.55 $\pm$ 0.05 eV for the state located
at 15.45 $\pm$ 0.10 eV and 1.35 $\pm$ 0.25 eV for
the state located at 16.92 $\pm$ 0.10 eV. These correspond to lifetimes 
of 1.20 $\pm$ 0.10 fs and 0.49 $\pm$ 0.08 fs, respectively. 
Both these values are larger than that of the corresponding atomic oxygen case 
which has an observed lifetime of 0.30 $\pm$ 0.02 fs for the $2s2p^5~^1P^o$ state.

As already indicated by the Fano profile indices $q$ of the resonances 
located at 15.45 $\pm$ 0.10 eV and 16.92 $\pm$ 0.10 eV (see Figure \ref{1D} (a)), 
with lifetimes of 3.85 $\pm$ 0.42 fs and 2.15 $\pm$ 00.15 fs respectively.
Both these states have lower values than the atomic oxygen case of 4.25 $\pm$ 0.80 fs. 
Accordingly, in the case of the smaller profile, the index $q$ for 
sulfur indicates a less efficient coupling.
At a photon energy of 16.92 $\pm$ 0.10 eV a broad autoionization 
resonance is observed with an absolute photoionization cross section 
of 23.6 $\pm$ 2.5 Mb and a shoulder at 15.45 $\pm$ 0.10 eV with a
cross section value of 9.0 $\pm$ 1.2 Mb. The structures are due to 
$3s^23p^4~ ^1D \rightarrow 3s3p^4 np~^1P^o,^1D^o,^1F^o$
transitions with $n > 3$.  The fitted experimental data
yield a Fano profile $q$ value of 2.15 $\pm$ 0.15. 

We note for the case of the $3s^23p^4~ ^1S$ metastable state, the $3s^23p^3(^2P^o)5s~^1P^o$ resonance state
lies below the S$^+(^2D^o)$ threshold and is therefore an interloping resonance \cite{qb3},
as illustrated in Figure \ref{thres1S-theory}, which disrupts the regular Rydberg pattern. Above the S$^+(^2D^o)$ threshold 
region and below the S$^+(^2P^o)$ threshold the $3s^23p^3(^2P^o)ns~^1P^o$ series follows it normal Rydberg pattern 
as no interlopers are present.  Only the lowest member $3s^23p^3(^2P^o)3d~^1P^o$ of the 
$3s^23p^3(^2P^o)nd~^1P^o$ resonance series was found. Our theoretical values for the positions and quantum defects 
are seen from Table \ref{meta-1S} to be in suitable agreement with the measurements of Yang and co-workers \cite{Jackson2008a}. 
The $3s^23p^3(^2P^o)3d~^1S^o$ resonance found by Yang and 
co-workers \cite{Jackson2008a}, located at 9.2035 eV, is forbidden by dipole selection rules in $LS$-coupling 
but shows up as a very faint resonance feature at approximately 9.2096 eV in our theoretical cross section.

\section{Discussion}
Photodissociation of \ce{CS2} at an excitation wavelength
of 193 nm   {(6.424 eV)} with subsequent photoionization of the neutral, formed
photofragments in the photon energy range between 9.25 and 30 eV is reported.
For atomic sulfur formed in the electronic states S$(3s^23p^4~^3P)$ and S$(3s^23p^4~^1D)$ 
measurements are performed up to 30 eV for both states.
Autoionizing Rydberg resonances in the energy below 13.40 eV 
are observed and a detailed analysis is carried out for them.

Subtracting the proportion of S$(3s^23p^4~^3P)$ from the photo-ion yield curves above
that of S$^+$ from S$(3s^23p^4~^3P)$ and  S$(3s^23p^4~^1D)$, provides atomic
triplet sulfur S$(^3P)$ from the photodissociation of \ce{S2O} 
for photon energies in the range 10.25 eV to 30 eV.
This allows photoionization measurements to be made on the triplet 
state of atomic sulfur for photon energies over this photon energy range.
For S$(3s^23p^4~^3P)$, autoionization resonances occurring in 
the experimental and theoretical investigations, i.e. in the photon energy range
between 12 -- 25 eV, intense window 
resonances are observed, that converge to the fourth ionization at 20.20 eV, as 
shown in Figure \ref{3P} and Figure \ref{1D} 
and illustrated more vividly in Figure \ref{window}.

Similarly, the photoion yield curve of electronically excited atomic
sulfur S$(3s^23p^4~^1D)$ is determined by subtracting 
the amounts of S$(3s^23p^4~^3P)$ from the S$^+$ yield 
curves determined in the pump-probe experiments
in the energy regime up to 30 eV for \ce{CS2} molecule. 
Here, the first few members of the autoionizing Rydberg resonances series below
the $^2D^o$ threshold (located at 11.06 eV) and the $^2P^o$ threshold (located 12.256 eV) are observed.
At a photon energy of 16.92 $\pm$ 0.10 eV, a highly intense and a strongly broadened autoionization resonance is found. 
This resonance has a shoulder at 15.45 $\pm$ 0.10 eV and may be due to 3s-inner shell excitation. 
The large Auger width of 1.35 $\pm$ 0.25 eV for this resonance dominates
the photoionization spectra of the electronically excited S$ (3s^23p^4~^1D)$ in this photon energy region. 
Over the photon energy range 16 -- 20 eV the cross section 
is dominated by remnants of the molecular precursors from the photolysis process. 
The origin of these processes is discussed in Section IV. A.
This enhancement is not observed in the corresponding calculated atomic photoionization cross section 
so we conclude they are due to molecular effects.

\section{Summary}
The photoionization cross sections for atomic sulfur in the states S$(3s^23p^4~^3P)$ and S$(3s^23p^4~^1D)$, 
are an important basis for atmospheric and astrophysical models. Broad autoionizing resonances,
observed in the case of electronically excited sulfur atoms are found by pump-probe experiments starting from molecular precursors. 
Numerous autoionization resonances are observed which facilitate the assignment of the sulfur species under study. 
Specifically, experimental measurements and large-scale photoionization cross sections calculations are presented
for the ground and metastable states of atomic sulfur. A detailed analysis has been performed 
for the resonance features observed in the corresponding cross sections and the Rydberg series 
assigned spectroscopically. Below 13.5 eV similar resonance features are seen in the 
experimental measurements and the theoretical work, which have been analyzed and compared. 
Overall the resonance features observed in this photon energy region show
suitable agreement between experiment and theory.

For the photon energy range 16 -- 20 eV the experimental cross section 
is dominated by the remnants of the molecular precursor from the photolysis process not 
present in the theoretical calculations for this atomic species. The possible origin of these processes is discussed.
%
%
%
%
\begin{acknowledgments}
Financial support by the German Research Foundation (DFG) is gratefully acknowledged (RU 420/7-1). 
B. M. McL. acknowledges support by the US National Science Foundation under the visitors program through a grant to ITAMP
at the Harvard-Smithsonian Center for Astrophysics, Queen's University Belfast through a visiting research fellowship (VRF) and
the hospitality of ER and the Physikalische Chemie Department of the Freie Universit\"{a}t of Berlin, during a recent research visit. 
This research used resources of the National Energy Research Scientific Computing Center, 
which is supported by the Office of Science of the U.S. Department 
of Energy (DOE) under Contract No. DE-AC02-05CH11231. 
The computational work was performed at the National Energy Research Scientific
Computing Center in Oakland, CA, USA 
and at The High Performance Computing Center Stuttgart (HLRS) 
of the University of Stuttgart, Stuttgart, Germany. 
This research also used resources of the Oak Ridge Leadership Computing Facility 
at the Oak Ridge National Laboratory, which is supported by the Office of Science 
of the U.S. Department of Energy (DOE) under Contract No. DE-AC05-00OR22725.
\end{acknowledgments}
%
%
%
%
\bibliographystyle{apsrev4-1}
\bibliography{final}

\end{document}